\documentclass[useAMS,usenatbib]{mn2e}
\usepackage{graphicx}
\usepackage{longtable} 
\usepackage{deluxetable}
\usepackage{lscape}
\usepackage{amsmath}

\usepackage{fancyvrb}
\usepackage{color}
\usepackage[ascii]{inputenc}

\graphicspath{{figs/}}

\def\apjl{ApJL }
\def\aj{AJ }
\def\apj{ApJ }

\def\apjs{ApJS }
\def\araa{ARA\&A }
\def\aap{A\&A }

\def\mnras{MNRAS }

\def\nar{New Astron. }

\def\halpha{\mbox{H$\rm \alpha$}}
\def\sloanr{\mbox{r'}}
\def\sloani{\mbox{i'}}
\def\ri{\mbox{r'-i'}}
\def\rha{\mbox{r'-$\rm {H\alpha}$}}
\def\iJ{\mbox{i'-J}}
\def\J{\mbox{J}}

\def\er{\mbox{$\rm \sigma_{r'}$}}
\def\eha{\mbox{$\rm \sigma_{H\alpha}$}}
\def\ei{\mbox{$\rm\sigma_{i'}$}}
\def\ej{\mbox{$\rm \sigma_{J}$}}

\def\param{\mbox{\boldmath$\theta$}}
\def\data{\mbox{\boldmath$d$}}
\def\ewha{\mbox{EW$\rm _{H\alpha}$}}
\def\lacc{\mbox{L$\rm _{acc}$}}
\def\lha{\mbox{$\rm L_{H\alpha}$}}
\def\macc{\mbox{$\rm \dot M_{acc}$}}
\def\mass{\mbox{$\rm M_*$}}
\def\age{\mbox{$\rm \tau$}}
\def\radius{\mbox{$\rm R_*$}}

\def\rin{\mbox{R$\rm _{in}$}}
\def\av{\mbox{A$\rm _{V}$}}
\def\a0{\mbox{A$\rm _{0}$}}
\def\sedobs{\mbox{SED$\rm _{obs}$}}
\def\sedint{\mbox{SED$\rm _{int}$}}
\def\sedapp{\mbox{SED$\rm _{app}$}}
\def\lhacont{\mbox{L$\rm (H\alpha)_{cont}$}}

\def\msol{\mbox{$\rm M_\odot$}}
\def\rsol{\mbox{$\rm R_\odot$}}
\def\lsol{\mbox{$\rm L_\odot$}}

\def\sqdeg{\mbox{${\rm deg}^2$}}

\def\lesssim{\mathrel{\hbox{\rlap{\hbox{\lower4pt\hbox{$\sim$}}}\hbox{$<$}}}}
\def\gtrsim{\mathrel{\hbox{\rlap{\hbox{\lower4pt\hbox{$\sim$}}}\hbox{$>$}}}}

\begin{document}
\title[Bayesian inference of T Tauri star properties]{Bayesian inference of T Tauri star properties using multi-wavelength survey photometry}
\author[Barentsen et al.]
{Geert Barentsen$^{1,2}$\thanks{E-mail: geert@barentsen.be}, J. S. Vink$^1$, J. E. Drew$^2$, S. E. Sale$^3$ \\
$^1$Armagh Observatory, College Hill, Armagh BT61 9DG, U.K.\\
$^2$Centre for Astrophysics Research, Science and Technology Research Institute, University of Hertfordshire, Hatfield AL10 9AB, U.K.\\
$^3$Rudolf Peierls Centre for Theoretical Physics, Keble Road, Oxford OX1 3NP, U.K.\\
}

\date{Received 2012 April 15; Accepted 2012 November 21.}

\pagerange{\pageref{firstpage}--\pageref{lastpage}} \pubyear{2012}

\maketitle

\label{firstpage}
\begin{abstract}
There are many pertinent open issues in the area of star and planet formation. Large statistical samples of young stars across star-forming regions are needed to trigger a breakthrough in our understanding, but most optical studies are based on a wide variety of spectrographs and analysis methods, which introduces large biases. 

Here we show how graphical Bayesian networks can be employed to construct a hierarchical probabilistic model which allows pre-main sequence ages, masses, accretion rates, and extinctions to be estimated using two widely available photometric survey databases (IPHAS \sloanr/\halpha/\sloani\ and 2MASS J-band magnitudes). Because our approach does not rely on spectroscopy, it can easily be applied to homogeneously study the large number of clusters for which Gaia will yield membership lists. 

We explain how the analysis is carried out using the Markov Chain Monte Carlo (MCMC) method and provide Python source code. We then demonstrate its use on 587 known low-mass members of the star-forming region NGC\,2264 (Cone Nebula), arriving at a median age of 3.0\,Myr, an accretion fraction of $20\pm2\%$ and a median accretion rate of $10^{-8.4}$\,\msol/yr.

The Bayesian analysis formulated in this work delivers results which are in agreement with spectroscopic studies already in the literature, but achieves this with great efficiency by depending only on photometry. It is a significant step forward from previous photometric studies, because the probabilistic approach ensures that nuisance parameters, such as extinction and distance, are fully included in the analysis with a clear picture on any degeneracies.

\end{abstract}


\section{Introduction}
Large uncertainties remain with respect to the mechanisms and timescales of star and planet formation. While it has been established that young solar-like stars stop accreting and lose their protoplanetary discs on a timescale of $\sim$1 to 10~Myr \citep{haisch2001,fedele2010}, there is no full understanding of the interplay between the various physical mechanisms which affect disc evolution \citep{williams2011}.

Large statistical samples of young stars are needed to make the breakthrough in our understanding. Whilst such samples have recently become available through infrared photometry which traces circumstellar {\it dust} \citep[][]{evans2009}, they are not available for emission-line studies which traces material in the {\it gas} phase. Understanding the evolution of the gas with respect to the dust is of critical importance for testing competing models of disc evolution and planet formation \citep[e.g.][]{najita2007,owen2011,espaillat2012}.

Existing gas emission-line studies have mostly relied on spectroscopy, which could only be obtained for limited numbers of stars in nearby star-forming regions \citep[e.g.][]{gullbring1998,natta2004,herczeg2008}. Moreover, it is often hard to inter-compare the results from different regions, because they have been obtained using a variety of spectrographs and analysis methods which complicate the analysis.

The increasing availability of data from large photometric surveys allows samples to be obtained across star-forming regions which are both larger and more homogeneous. For example, the INT Photometric \halpha\ Survey (IPHAS) covers 1800 deg$^2$ of the Northern Galactic Plane using \sloanr/\sloani\ broad-band and \halpha\ narrow-band filters \citep{drew2005,idr}. This survey is particularly relevant to star formation studies, because \halpha\ photometry allows a statistical appraisal of gas accretion rates to be made for massive clusters at large distances \citep{demarchi2010,spezzi2012}.

So far, we have used the IPHAS survey to study pre-main sequence stars in just one star-forming region: IC\,1396 in \citet{barentsen2011}, hereafter Paper~I. We used H$\alpha$ narrow-band photometry to identify T Tauri stars and estimate their accretion rates, whilst simultaneously estimating ages and masses from the (\ri)/\sloanr\ colour-magnitude plane.

Before we can apply our methodology to the entire IPHAS (and future VPHAS) Galactic Plane region, we first need to develop more powerful tools. 
Whilst the results of Paper~I were in good agreement with independent spectroscopic measurements, the method suffered from the drawback that a fixed amount of extinction was assumed for all objects, as there is no straightforward method to obtain this parameter simultaneously with ages, masses and accretion rates from colour-magnitude or colour-colour diagrams. The increasing sophistication of models and the so-called ``data deluge'' from surveys requires more powerful inference tools to be adopted, as there are limitations to the information content to be extracted from two-dimensional diagrams. 

The generic mathematical solution to the problem of understanding which parameter-space regions match a set of observations is called \emph{Bayesian inference}. The theoretical principles of the method have been understood for decades, but the widespread adoption in astrophysics has only taken off in recent years owing to the advances in both algorithms and computing power. So far, the Bayesian framework has become the favoured tool for e.g. the determination of cosmological parameters \citep[][]{trotta2008}, the analysis of transit light curves \citep[][]{ford2005,kipping2012} or the determination of meteor rates \citep{barentsen2011b}. The approach has also been explored for estimating ages and extinctions for main-sequence stars \citep{pont2004,jorgensen2005,bailerjones2011}. In the context of star formation, the method has recently been used to perform a dynamical membership analysis of the Sco OB2 association \citep{rizzuto2011} and to assess the accuracy of pre-main sequence models \citep{gennaro2012}. However, the method has so far not been used to tackle the common problem of estimating the basic parameters of individual pre-main sequence stars.

In this paper we show how Bayesian inference can be used to simultaneously determine extinction, stellar ages \& masses and accretion rates for known members of a star-forming region. We will demonstrate the method on NGC\,2264, which is one of the best-studied regions within the IPHAS survey area and has a very complete membership list \citep{handbook2264,sung2008}. We note that the future Gaia astrometric survey is expected to yield accurate membership lists for hundreds of clusters \citep{bailer2009}, therefore our strategy to re-analyse a large sample of known cluster members in a homogeneous way is likely to become an increasingly important tool.

In \S2 we motivate our approach and specify the model. In \S3 we explain the application to NGC\,2264 and in \S4 we present the results. In \S5-\S7 we discuss the outcome, present future extensions and summarise the conclusions.

\section{Method: Bayesian Inference}
\label{method}
Our aim is to determine ages, masses and accretion rates from IPHAS \sloanr/\sloani/\halpha\ magnitudes, while simultaneously constraining the extinction by adding 2MASS J-band magnitudes to the dataset. We employ Bayesian Inference for this purpose. In \S 2.1 we describe the motivation for the method, in \S 2.2-2.3 we explain the formalisms and implementation, while in \S 2.4-2.5 we explain the practical use.

\subsection{Motivation}

\subsubsection{Characteristics of T Tauri stars}

In the current picture of star formation, solar-like stars are thought to assemble the majority of their mass during the first few $10^5$ years after the initial collapse of their parent molecular cloud \citep{evans2009}. Within $\sim$1\,Myr the envelope of gas and dust clears and the newly formed stars become visible at optical wavelengths, from which point they are commonly called \emph{T Tauri} stars. These objects continue to grow by accreting material from a circumstellar accretion disc, which does not exceed 10-20\% of the stellar mass and is dispersed within a few million years \citep{hartmannbook,haisch2001,fedele2010}.

Mass accretion is thought to take place along magnetic field lines which connect the disc to the star. Infalling gas is essentially on a ballistic trajectory, falling on to the stellar surface at near free-fall velocities, thereby producing hot impact shocks which generate excess UV and optical continuum emission \citep{calvet1998,gullbring2000}. The accretion energy released in these shocks also heats the infalling gas, which in turn produces strong \halpha\ emission. 

\begin{figure}
\includegraphics[width=\linewidth]{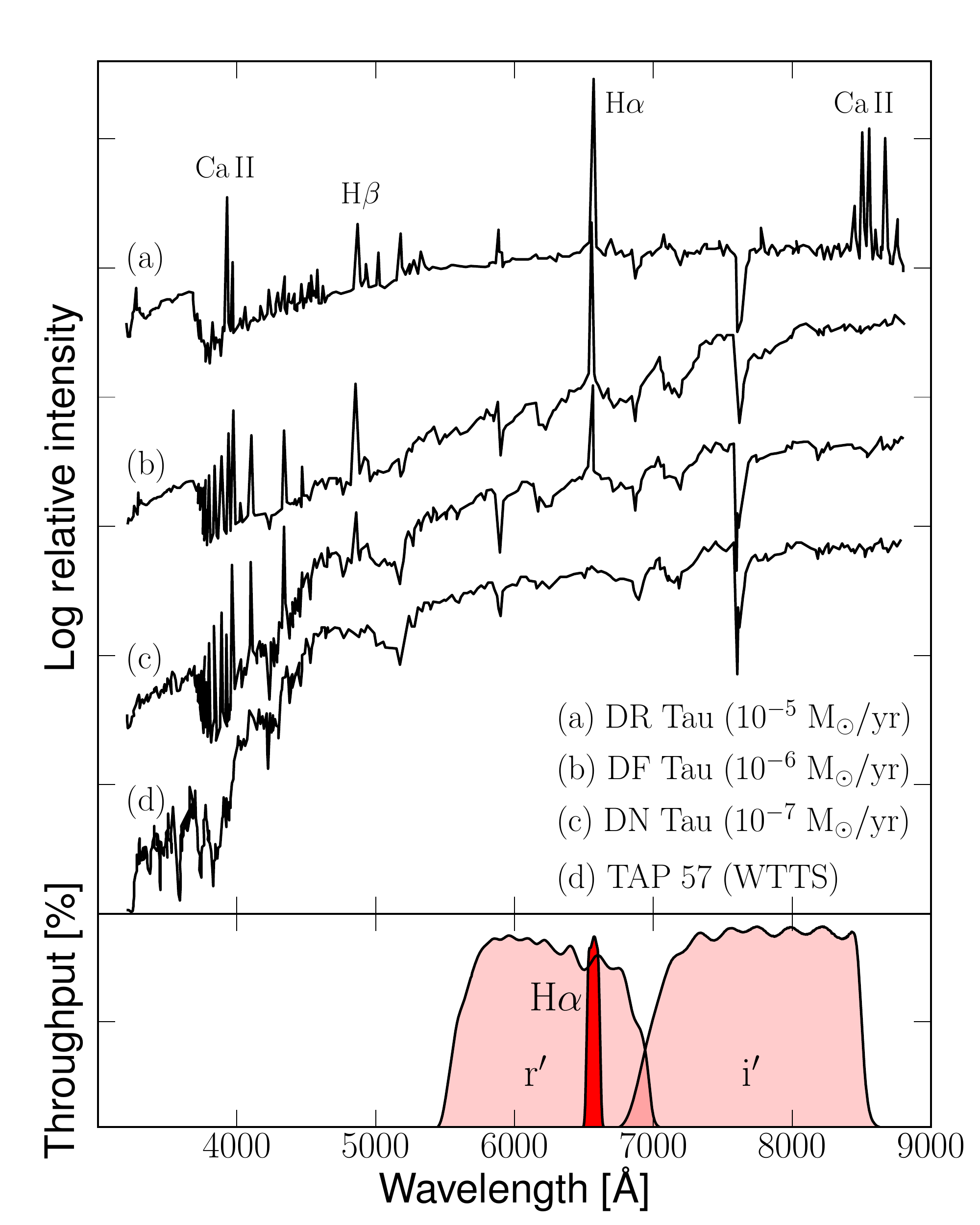}
\caption{Top panel: spectra of four T Tauri stars of similar spectral type (late K) as published by \citet{bertout1989} shown in order of decreasing accretion rates as determined by \citet{hartigan1995}. The most notable features are (i) blue continuum excess and line veiling below $\sim$6000\,\AA; (ii) hydrogen Balmer lines in emission; (iii) Ca\,II H/K and Infrared Triplet (IRT) lines in emission. The bottom spectrum is a weak-lined T Tauri star (WTTS) which does not show signs of ongoing mass accretion. Bottom panel: filter transmission curves for the IPHAS \sloanr/\halpha/\sloani\ filters. The 2MASS J-band filter is located beyond the range of the spectra near 12\,000\,\AA.}
\label{fig:bertout} 
\end{figure}

This is illustrated in Fig.~\ref{fig:bertout}, where we show literature spectra of four T Tauri stars of a similar spectral type (late K), shown in order of accretion rates which have been estimated from the UV/optical continuum excess emission. The spectra illustrate that the \halpha\ line strength is correlated with the blue excess, both thought to be a result of the release of accretion energy. In contrast, the \sloanr\ and \sloani\ bands appear least affected by accretion and are thus the most appropriate tracers for stellar age and mass. Hence, in our past study of IC\,1396 (Paper~I) we used the (\ri)/\sloanr\ plane to estimate ages and masses from model isochrones, whilst using the (\ri)/(\rha)\ plane to estimate \halpha\ equivalent widths and accretion rates.

We note that at exceptionally high accretion rates (${\gtrsim10^{-6}}$\,\msol/yr) there is evidence for accretion-induced continuum veiling to occur in the \sloanr\ and \sloani\ bands, which may affect the age and mass estimates. The source of the excess emission at these wavelengths is not well understood at present \citep{fischer2011}. The effect is small and will be ignored in what follows, because our study includes only objects with lower accretion rates. We will return to this topic at the end of the paper however (\S\ref{sec:lowlikelihood}).

\subsubsection{Solving the degeneracy between age, mass, extinction and \halpha\ emission}
\label{sec:degeneracy}
\begin{figure}
\includegraphics[width=\linewidth]{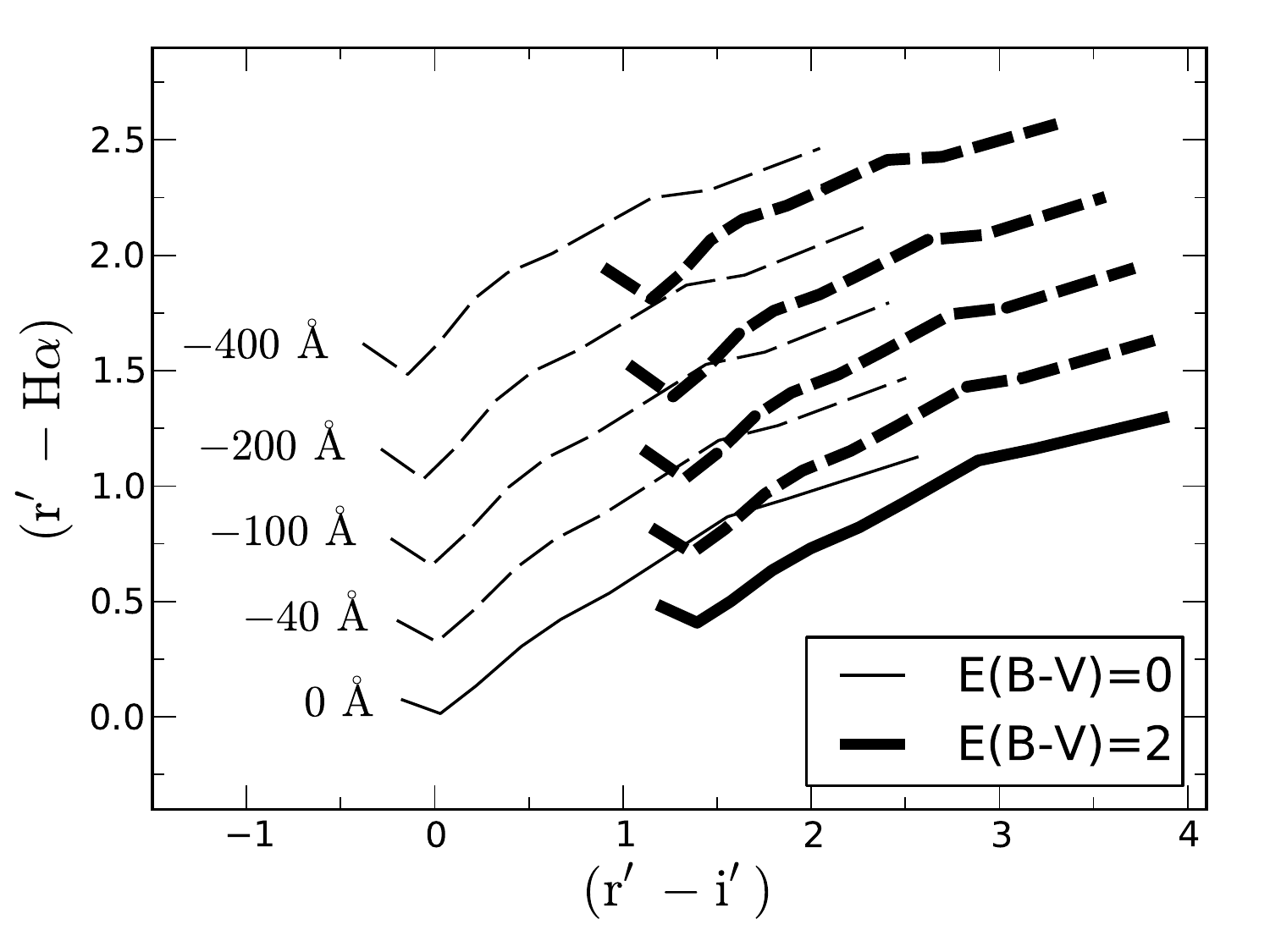}
\caption{Position of the unreddened and reddened main sequence (thin and thick lines) in the IPHAS (\ri)/(\rha)\ plane, shown from early O to late M-type stars. We also mark the position of the main sequence for increasing levels of \halpha\ emission (dashed lines, indicated by their \halpha\ EWs in units \AA). The figure illustrates the degeneracy between reddening, spectral type and \halpha\ emission in this plane. These simulations are taken from Paper~I.
}
\label{fig:ccd_iphas}
\end{figure}
The results obtained in Paper~I were in good agreement with spectroscopic measurements from the literature. However, the method suffered from the drawback that a fixed amount of extinction was assumed for all objects, because optical colour-colour diagrams show a well-known degeneracy between reddened high-mass (early-type) stars and unreddened low-mass (late-type) stars. This may surprise some readers, because the (\ri)/(\rha) plane is known for being able to break this degeneracy due to the strong difference in the \halpha\ line strength between early- and late-type stars \citep{drew2008}. However, this property cannot be exploited when \halpha\ is in emission. 

The degeneracy is illustrated in Fig.~\ref{fig:ccd_iphas}, where we plot the colour simulations from Paper~I to show the intrinsic position of the main sequence as a function of reddening (thick and thin solid lines). These solid curves do not overlap or intersect, i.e. there is a handle on the degeneracy. However, when increasing levels of \halpha\ emission are added (dashed lines), reddened early-type stars with \halpha\ in emission become entangled with unreddened late-type stars.

To resolve this degeneracy, we require additional colour(s) to be added to the IPHAS dataset. Our best-available option is to add the near-infrared J-band magnitude from the 2MASS survey \citep{2mass}, because the (\ri)/(\iJ) plane can be shown to break the degeneracy between reddening and spectral type for K- and M-type objects, which constitute the vast majority of T Tauri stars \citep{martin1997}. At the same time, the J-band is not commonly affected by excess emission from a circumstellar disc, unlike the 2MASS H- and K-bands at longer wavelengths \citep[][]{meyer1997}.

However, adding the J-band magnitude to our sample does not allow us to simply read the extinction for every object from the (\ri)/(\iJ) diagram straight away, as these colours do not depend on mass and extinction alone. For example, the \halpha-line falls inside the \sloanr-band, such that objects with \halpha\ in emission show an \sloanr-band excess ranging between -0.05 mag (\ewha $\cong$ -50\,\AA) and -1.0 mag (\ewha $\cong$ -1000\,\AA), albeit depending on the spectral type (Paper~I). In addition, the colours also depend on the stellar age as pre-main sequence stars tend to rise in effective temperature as they approach the main sequence.

Whilst adding the J-band data should offer us sufficient information to constrain the four parameters of interest, the problem remains how these constraints can be inferred in practice?

\subsubsection{Inferring parameters from photometry}

The traditional method used to estimate stellar parameters from photometry is to place the observed objects in two-dimensional colour/magnitude diagrams, together with the output of evolutionary models. Such an approach is useful when the number of free parameters is small. However, when the number of dimensions in the parameter or observable space increases, there is no obvious way to decide which plane/parameter-combinations should be employed. For example, our \sloanr/\halpha/\sloani/J dataset can be used to construct no less than 15 different colour-colour diagrams and 24 colour-magnitude diagrams, all of which depend to some extent on all of our parameters of interest.

We could devise an ad-hoc algorithm to make use of all the available information, for example: we could iteratively fit model parameters in multiple planes until a certain convergence criterion is met. We can also employ a range of data modelling methods to help us link observations to parameters (e.g. \emph{Neural Networks}, \emph{Principal Component Analysis}). However, it is unclear how the results obtained by such methods depend on ad-hoc choices made in the design of the algorithms (e.g. the structure of the Neural Network). Moreover, there is no clear way for these methods to obtain meaningful error bars which take full account of all known sources of uncertainty \citep{ford2005}.

A better approach is to obtain \emph{maximum likelihood} estimates using \emph{expectation-maximisation} (EM) algorithms. These methods require the user to define a predictive model which estimates the likelihood of an observation given a set of model parameters. The model is then optimised to find the set of parameters which are \emph{most likely} to explain the observed data (in the special case of a Gaussian likelihood model this is \emph{$\chi^2$-fitting}). 

While EM presents a significant improvement over ad-hoc methods, it suffers from the drawback that only a single ``best-fit'' point estimate is provided, regardless of whether or not a unique solution exists. This is a particular concern in astronomy, where model uncertainties and data sparsity imply that there is often a ``family'' of likely solutions which occupy degenerate or multi-modal regions in the parameter space. This problem is often solved by keeping one or more of the degenerate ``nuisance'' parameters fixed, but such assumptions invariably reduce the ability of the model to capture reality.

Rather than focusing on finding a ``best-fit'' estimate, it is better to employ the predictive model to infer the full probability distribution for all possible solutions. This approach is called \emph{Bayesian inference} (though we note that some authors prefer the term \emph{probabilistic inference} to distinguish the approach from best-fit EM methods which also employ the Bayes' theorem.) 

Obtaining the full distribution may seem impractical at first sight, because point estimates are useful for plotting and tabulation. However, a full distribution can be reduced to a point estimate by computing the expectation value or median, which, unlike the maximum likelihood, take full account of the distribution (that is, an expectation value minimises the variance, while the median minimises the mean absolute error.) Moreover, knowledge of the full distribution allows meaningful confidence intervals and covariances to be quantified and visualised, by marginalising over the nuisance parameters.

In what follows we provide a formal description of the method and explain how it is applied to our problem.

\subsection{Solution: Bayesian inference}
\subsubsection{Formalism}
The basic idea is to create a parameterised model which is able to reproduce the data and its uncertainty, and then compare that model for different sets of parameters against the observations in a probabilistic way.

Let $\param  = \{\theta_1,\ldots, \theta_n\}$ represent a set of unknown model parameters and let $\data = \{d_1,\ldots, d_i\}$ represent a set of observed data. We can construct a \emph{likelihood model} $P(\data | \param)$ which computes the probability for an observation to occur under a given a set of parameters. Such a model can be computed using the best-available knowledge, and is limited only by scientific complexity and computing power.

Of course our aim is not to understand which observations are expected given the parameters, but inversely, \emph{which of the various possible sets of parameters best explain a given observation}. This is expressed by the function $P(\param | \data)$, called the \emph{posterior} distribution, which can be linked to the likelihood model using the \emph{chain rule} from probability theory:
\begin{equation}
P(\param | \data) \cdot P(\data) = P(\param, \data) = P(\data | \param) \cdot P(\param),
\end{equation}
which leads to the well-known theorem by Bayes:
\begin{equation}
P(\param | \data) = \frac{ P(\data | \param) \cdot P(\param) }{ P(\data) },
\end{equation}
where $P(\param)$ is called the {\em prior}, which encodes any a priori knowledge about the parameters which we wish to include in our model (including the allowed physical bounds). The denominator, $P(\data)$, may be thought of as a normalising constant which does not effect the shape of the posterior distribution and can be ignored, i.e.:
\begin{equation}
P(\param | \data) \propto P(\data | \param) \cdot P(\param),
\end{equation}
or simply:
\begin{equation}
P(\param | \data) \propto P(\param, \data).
\end{equation}
Thus, the key to identify the regions in the parameter space which explain a set of observations is the ability to compute the joint probability distribution $P(\param, \data)$.

\subsubsection{Constructing the joint distribution}
$P(\param, \data)$ can be thought of as the model which defines how the theoretical parameters and observations relate to each other. We now explain how this model is formulated for our application. 

First, let us define the set of unknown variables \param\ and the set of observables \data. 
The free parameters of principal interest in our work are mass (\mass), age (\age), mass accretion rate (\macc) and extinction (\a0).
For syntactic convenience, we add a set of additional variables which help us formulate the model, such as a star's intrinsic spectral energy distribution (\sedint). Their meaning is explained in Table~\ref{notation}:
\begin{align}
\param = { } & \{\mass,\ \age,\ \macc,\ \a0,\ \sedint,\ \rin,\ \lha,\ \\
&\ \ewha,\ d,\ \sedapp\}.
\end{align}
We note that the additional variables are either determined by the four free parameters of interest, or they are so-called \emph{nuisance parameters} which will be constrained by a strong prior assumption (e.g. in what follows the distance $d$ will be constrained by a strong prior based on literature estimates).

For the observed data of a star, we adopt the SED as characterised by the four apparent magnitudes and their uncertainties:  
\begin{equation}
\data = \sedobs = \{\sloanr,\ \halpha,\ \sloani,\ \J,\ \er,\ \eha,\ \ei,\ \ej\}. 
\end{equation}

Having defined \param\ and \data, we now formulate $P(\param,\data)$. This is a complex distribution with a high number of dimensions. It would be tedious to construct a single function which computes its value for all possible combinations of \param\ and \data.
We can greatly reduce the complexity however by exploiting the fact that many of the variables can be assumed to have (conditional) independence relationships.

\begin{table}
  \begin{tabular}{ l l }
  \hline
  \mass & Stellar mass (\msol)\\
  \age & Stellar age (Myr) \\
  \macc & Accretion rate (\msol\ $\mathrm{yr^{-1}}$) \\
  \rin & Inner disc truncation radius (\radius) \\
  \sedint & Modelled intrinsic SED $\rm \{M_{\sloanr},\ M_{\halpha},\ M_{\sloani},\ M_{\J} \}$ \\
  \lha & Excess \halpha\ luminosity (\lsol) \\
  \ewha & \halpha\ emission Equivalent Width (\AA) \\
   d & Distance (pc) \\
   \a0 & Extinction parameter (mag) \\
   \sedapp & Modelled apparent SED $\rm \{m_{\sloanr},\ m_{\halpha},\ m_{\sloani},\ m_{\J} \}$ \\
  \sedobs & Observed SED $\rm \{\sloanr,\ \halpha,\ \sloani,\ \J,\ \er,\ \eha,\ \ei,\ \ej\}$ \\
  \hline
  \end{tabular}
  \caption{Notation.}
  \label{notation}
\end{table}
\begin{figure}
\includegraphics[width=\linewidth]{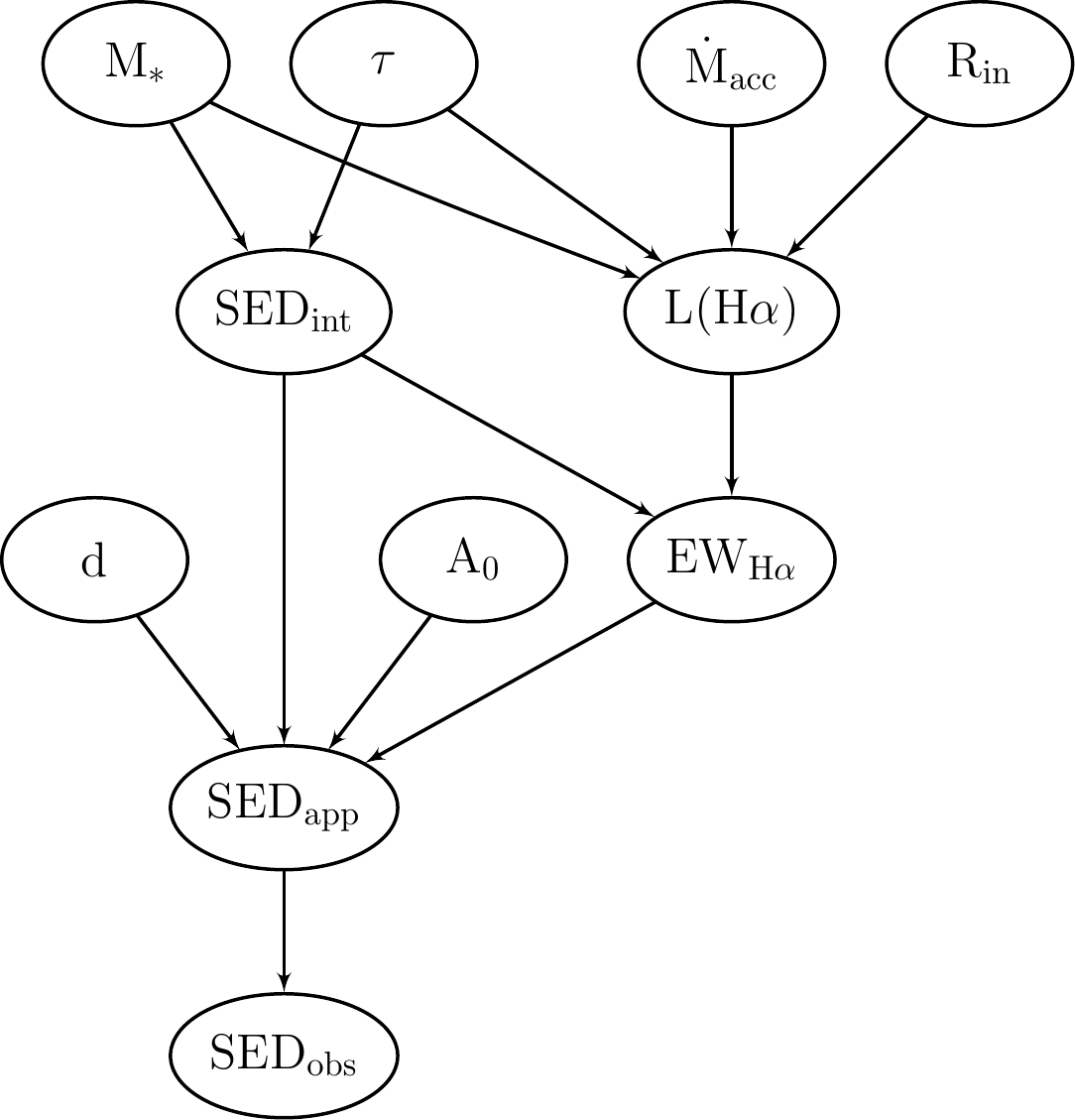}
\caption{Bayesian network representing the dependencies between the variables in our inference model. Nodes which are not directly connected represent variables which are conditionally independent of each other given their parents.}
\label{dag} 
\end{figure}

A convenient and concise way to represent the dependencies amongst variables is to use a \emph{directed acyclic graph} (i.e. a graph with no loops), often referred to as a \emph{probabilistic graphical model} or a \emph{Bayesian network}. These terms are synonyms for a syntactic method whereby variables are represented by nodes in a graph, and dependencies between those variables are represented by arrows. Intuitively, an arrow from variable $A$ to $B$ indicates that $A$ has a \emph{direct influence} on $B$.
Formally, the Bayesian Network is constructed as follows: \citep{aibook} 
\begin{enumerate}
\item each node corresponds to a variable from \param\ or \data;
\item a set of arrows connect pairs of nodes. If there is an arrow from node $A$ to node $B$, then $A$ is said to be a \emph{parent} of $B$. Pairs of nodes which are \emph{not} directly connected by an arrow represent variables which are assumed to be conditionally independent of each other given their parents;
\item the graph should not have directed cycles, this would be a sign of recursive reasoning.
\end{enumerate}

The Bayesian network for our model is shown in Fig.~\ref{dag}. For example, the graph shows arrows pointing from \mass\ and \age\ to the intrinsic SED, which represents our assumption that we only need to know \mass\ and \age\ to infer \sedint, i.e.:
\begin{equation}
P(\sedint\ |\ \param,\data) = P(\sedint\ |\ \mass,\age)
\end{equation}
We may use this property to simplify the formulation of $P(\param,\data)$ as follows. Let $(n_1, \dots, n_k)$ be the set of all variables. The chain rule allows us to write its joint distribution as a product of conditional probabilities:
\begin{equation}
P(n_1, \dots, n_k) = \prod_{i=1}^k P(n_i\ |\ n_{i+1}, \dots , n_k).
\label{eqv}
\end{equation}
Given the specification of a Bayesian network over all nodes $n_i$, Eqn.~\ref{eqv} may be simplified using the property of conditional independence:
\begin{equation}
P(n_1, \dots, n_k) = \prod_{i=1}^k P(n_i\ |\ Parents(n_i)),
\end{equation}
where $Parents(n_i)$ is the set of nodes which have an arrow pointing to $n_i$ in the network, and $Parents(n_i) \subseteq \{n_{i+1}, \dots , n_k\}$. The last condition can always be satisfied by pre-ordering the nodes in a way that is consistent with the partial order given by the network. 

$P(\param,\data)$ can now be written as the product of a series of lower-dimensional distributions:
\begin{align}
P(\param,\data)  = { } & \prod_{i=1}^k P(n_i\ |\ Parents(n_i)) \label{model} \\
 = { } & P( \mass ) \cdot P( \age ) \cdot P( \macc ) \label{priors1} \\
& \cdot\, P( \rin ) \cdot P( d ) \cdot P( \a0 ) \label{priors2} \\
& \cdot\, P( \sedint\ |\ \mass, \age ) \label{psedint} \\
& \cdot\, P( \lha\ |\ \mass, \age, \macc, \rin) \label{plhaexc} \\
& \cdot\, P( \ewha\ |\ \lha, \sedint ) \label{pew} \\
& \cdot\, P( \sedapp\ |\ \sedint, \ewha, d, \a0) \label{psedapp} \\
& \cdot\, P( \sedobs \ |\ \sedapp ) \label{psedobs}
\end{align}
It is trivial to see that rewriting the equation in this way makes it far easier to define our probabilistic model, which is now a combination of small hierarchical ``sub-models''.  The models for parameters which do not have parents are called the \emph{priors} (Eqns.~\ref{priors1}-\ref{priors2}), whilst the others are called the \emph{likelihoods} (Eqns.~\ref{psedint}-\ref{psedobs}). In what follows we explain how each of the above factors are computed, which acts as a formal specification of our parameter inference model.

\subsection{Priors and Likelihoods}
\subsubsection{Priors}
\label{priors}
A challenge that comes with the use of Bayes' theorem is the choice of the priors. We should not shy away from this task however, as there is no inference without assumptions. Being forced to define them in a clear way can be seen as an advantage over other methods (e.g. $\chi^2$-fitting bears the implicit assumption of uniform priors on the free parameters, yet the consequence of this assumption is rarely considered.) In the case of our most influential prior, P(\a0), we will quantify the extent of its influence in \S\ref{sec:extinctionprior} at the end of the paper.

The prior distributions are summarised in Table~\ref{tab:priors} and explained as follows:

\begin{table}
  \begin{tabular}{ l l }
  \hline
  Prior & Distribution \\
  \hline
  P(\mass) & $\sim
\left\{
	\begin{array}{ll}
		\mass^{-1.3}  & \mbox{if\ \ } 0.1 < \mass < 0.5 \\
		0.5\,\mass^{-2.3} & \mbox{if\ \ }  0.5 \leq \mass < 7
	\end{array}
\right.$  \\
  P($\log$ \age) & $\sim \mathcal{U}(5, 8)$   \\
  P($\log$ \macc) & $\sim \mathcal{U}(-15, -2)$  \\
  P(\rin) & $\sim \mathcal{N}(5, \sigma=2) \hspace{0.7cm} (\rin > 1)$   \\
  P(d) & $\sim \mathcal{N}(760, \sigma=5)$  \\
  P($\log$ \a0) & $\sim \mathcal{N}(-0.27, \sigma=0.46)$   \\
  \hline
  \end{tabular}
  \caption{Summary of priors. $\mathcal{U}(min,max)$ denotes a Uniform distribution, while $\mathcal{N}(\mu,\sigma)$ denotes a Gaussian.}
  \label{tab:priors}
\end{table}

\begin{itemize}
\item{P(\mass):} the mass prior follows the Initial Mass Function (IMF) due to \citet{kroupa2001}, truncated between 0.1 and 7\,\msol\ which are the limits of the stellar evolutionary model that we adopt in what follows. Objects outside this range are either saturated or fall below the detection limit of the IPHAS survey and so these truncation limits do not affect our results, other than constraining the parameter space to a sensible domain.
\item{P(\age):} the age prior is assumed uniform in the logarithm and truncated between 0.1 and 100\,Myr, which again corresponds to the limits of the evolutionary model.
\item{P(\macc):} the accretion rate prior is assumed uniform in the logarithm  and is truncated between $10^{-15}$ and $10^{-2}$\,\msol\,yr$^{-1}$. This range entails all the accretion rates commonly reported in the literature and goes well below the typical detection limit of $\sim10^{-10}$\,\msol\,yr$^{-1}$ we found in Paper~I.
\item{P(\rin):} the disc truncation radius follows a Gaussian distribution with a mean of $5\,\radius$ and standard deviation $2\,\radius$. This is a commonly used assumption based on the typical co-rotation radius of T Tauri stars \citep{gullbring1998}.
\item{P($d$):} for the distance prior we adopt a Gaussian distribution centred on the widely cited distance towards NGC\,2264 of 760\,pc \citep{sung1997}.
The standard deviation of 5\,pc reflects the approximate diameter of the cluster and therefore the uncertainty in our results will reflect only the relative distance errors. We note that a recent distance estimate by \citet{baxter2009} puts the region at 913~pc and so the systematic error in the distance may be significantly larger than 5~pc. However, because we aim to investigate the relative properties of objects in the cluster, rather than systematic errors for the cluster as a whole, we opt not to model the absolute distance uncertainty here. (Note that the Gaia survey will remove most of this uncertainty in the future.)
\item{P(\a0):} for the extinction parameter we adopt the empirical distribution of extinction for 202 candidate members as determined by \citet{rebull2002} on the basis of moderate-resolution spectroscopy of low-mass objects in NGC\,2264. We found that their distribution can be well-approximated as a log-normal with mean $\log A_0 = -0.27$ ($A_0=0.54$) and a broad standard deviation $\sigma \log A_0 =0.46$.
\end{itemize}

\subsubsection{\sedint\ likelihood (Eqn.~\ref{psedint})}
\label{sec:sedint}
The intrinsic SED of a young star is predicted as a function of mass and age as follows. 

First, intrinsic broad-band magnitudes $\mathrm{M_{r'}}$, $\mathrm{M_{i'}}$ and $\mathrm{M_{J}}$ are interpolated from the evolutionary model due to \citet{siess} for solar metallicity ($Z=0.02$). Their model consists of 29 separate mass tracks from 0.1 to 7~\msol\ from which isochrones can be computed using an online tool\footnote{http://www.astro.ulb.ac.be/$\sim$siess/WWWTools/Isochrones}. We used this tool to download the model at 50 different ages ranging between 0.1~Myr and 100~Myr, i.e. we obtained a dense sampling of the model in 1450 discrete points as a function of mass and age ($=29\cdot50$ points).

The \citeauthor{siess} model provides intrinsic magnitudes on the basis of the empirical conversion tables presented by \citet{kenyon1995}, which provide intrinsic colours and bolometric corrections as a function of stellar effective temperature. These tables only provide a calibration for the Cousins photometric system however, so we had to convert the model $\mathrm{R_C}$/$\mathrm{I_C}$ magnitudes to IPHAS \sloanr/\sloani\ using the transformations given in Paper~I.

We then approximated the evolutionary model as a continuous function by fitting 1450 Radial Basis Functions (RBF) to the collection of discrete model points using the Python module {\sc scipy.interpolate.rbf} (cf. Appendix~A). The resulting set of basis functions provides us with an accurate and fast tool to interpolate values from the grid.

Having obtained intrinsic magnitudes as a function of age and mass, we then predict the narrow-band magnitude $\mathrm{M_{H\alpha}}$ using the grid of IPHAS colour simulations presented in Paper~I, where we determined the intrinsic colour (\rha)\ as a function of (\ri) on the basis of a library of observed spectra.

These steps provide us with a forward model of the intrinsic magnitudes as a function of age and mass, i.e. $f_{\rm SED}(\mass,\age) = \{M_{\sloanr},\ M_{\halpha},\ M_{\sloani},\ M_{\J} \}$. For simplicity, we assume here that this model predicts absolute magnitudes with zero uncertainty. In reality, there are known to be significant systematic differences between pre-main sequence models, ranging up to 2-4 Myr in age and 0.2 \msol\ in mass (Paper~I). However, the modelling and study of these systematic effects are beyond the scope of this work, and so we adopt a deterministic likelihood:
\begin{equation}
P( \sedint\ |\ \mass, \age ) =
\left\{
	\begin{array}{ll}
		1  & \mbox{if\ \ } \sedint = f_{\rm SED}(\mass,\age) \\
		0 & \mbox{if\ \ } \sedint \neq f_{\rm SED}(\mass,\age)
	\end{array}
\right.
\label{eqn:sedint}
\end{equation}

\subsubsection{\lha\ likelihood (Eqn.~\ref{plhaexc})}
\halpha\ is the strongest emission line due to accretion, and its intensity can be used to trace the accretion luminosity.

As explained in \S2, accretion is thought to take place along magnetic field lines which act as channels connecting the disc to the star from an inner disc truncation radius \rin. The infalling gas is essentially on a ballistic trajectory, falling on to the star at near free-fall velocities, producing a hot impact shock \citep{calvet1998,gullbring2000}.
The energy released in these shocks heats the infalling circumstellar gas, which explains the \halpha\ emission.

Under the assumption that the free-fall gravitational energy released in the impact accretion shock is reprocessed entirely in the accretion energy \lacc, we may write:
\begin{align}
\rm L_{\rm acc} = \frac{G \mass \macc}{\radius} (1 - \frac{\radius}{\rin}),
\end{align}
and therefore:
\begin{align}
\log \frac{\lacc}{\lsol} = { } & 7.496 + \log \frac{\mass}{\msol} + \log \frac{\macc}{\msol yr^{-1}} \nonumber \\
& - \log \frac{\radius}{\rsol} + \log(1 - \frac{\radius}{\rm R_{\rm in}} ),
\label{eqn:lacc}
\end{align}
where \radius\ is derived from a forward model $f_{\radius}(\mass,\age)$ obtained by interpolating the Siess et al. model in the same way as described previously.

The accretion luminosity \lacc\ has previously been found to relate to \lha\ as a power-law relationship \citep[e.g.][]{herczeg2008,demarchi2010}. In Paper~I (fig.~8) we presented a compilation of 107 objects from the literature for which both \lacc\ and \lha\ estimates are available, whereby \lacc\ has been derived from blue continuum excess measurements. We used the {\sc stats.lm} function in the {\sc R} statistical environment to determine the linear least squares regression:
\begin{align}
\log \lha = (0.64 \pm 0.04) \log \lacc - (2.12 \pm 0.08). 
\label{eqn:lha}
\end{align}
This empirical relationship shows a significant scatter however (rms~=~0.43), which is commonly assumed to be caused by a combination of physical effects (e.g. absorption by stellar winds, uncertain extinction corrections, emission not due to accretion). The scatter in this relationship dominates the uncertainty in the inferred accretion rates (Paper~I). We take this into account by modelling the likelihood function as a Log-normal distribution:
\begin{align}
P( \lha\ |\ \mass, \age, \macc, \rin ) \sim \log\mathcal{N}(\log \lha, \sigma=0.43),
\end{align}
where $\log \lha$ is obtained by combining Eqns.~\ref{eqn:lacc} \& \ref{eqn:lha}.

\subsubsection{\ewha\ likelihood (Eqn.~\ref{pew})}
To infer \ewha\ we need to predict the stellar continuum luminosity in the \halpha\ band:
\begin{equation}
\lhacont = \mathrm L_V(\halpha) \cdot 10^{-0.4 \cdot [ \mathrm{M_{H\alpha}} + 0.03 ] } 
\end{equation}
where $\mathrm{M_{H\alpha} \in \sedint}$ is the previously estimated intrinsic magnitude of the star and $\mathrm L_V(\halpha) = 0.316~\lsol$ is the luminosity of Vega in the IPHAS \halpha\ passband (Paper~I). We may then obtain the equivalent width from its definition:
\begin{equation}
\ewha = - RW \cdot \frac{\lha}{\lhacont},
\end{equation}
where $RW = 95$\,\AA\ is the rectangular width of the IPHAS \halpha\ filter.

The uncertainty associated with the conversion of \lha\ into \ewha\ can be assumed negligible and for simplicity we take the likelihood to be deterministic.

\subsubsection{\sedapp\ likelihood (Eqn.~\ref{psedapp})}
\label{sedapplikelihood}
The apparent SED is obtained by correcting the intrinsic SED for the effects of distance, extinction and \halpha\ emission in three steps.

First, the distance modulus $5\log(d)-5$ is added to each of the intrinsic magnitudes. Second, offsets for the \sloanr/\halpha/\sloani\ magnitudes are obtained as a function of \a0\ and \ewha\ by means of RBF-interpolation from the pre-computed grid of simulated photometry from Paper~I. Finally, the offset for the J magnitude for extinction is computed using the reddening law due to \citet{schlegel}. The J-band offset cannot be computed in the same way as the other bands, because the optical spectral library on which our grid of simulated colours is based does not extend far enough into the near-infrared.

Again, this likelihood is assumed deterministic.

\subsubsection{\sedobs\ likelihood (Eqn.~\ref{psedobs})}
Finally, we compute the likelihood of observing \sedobs\ when expecting \sedapp. Assuming normally distributed uncertainties on the observed magnitudes, the likelihood is given by a multivariate Gaussian:
\begin{equation}
P( \sedobs \ |\ \sedapp ) = \frac{1}{(2\pi)^{k/2}|\mathbf\Sigma|^{1/2}} e^{-D^2/2},
\end{equation}
where $\mathbf\Sigma$ is the covariance matrix of the magnitudes in \sedobs\ and $D^2$ is given by
\begin{equation}
D^2 = (\sedobs-\sedapp)^T{\mathbf\Sigma}^{-1}(\sedobs-\sedapp).
\end{equation}

We assume that the uncertainties between magnitudes in \sedobs\ are uncorrelated, in which case $\mathbf\Sigma$ is a diagonal matrix and $D^2$ can be simplified to
\begin{equation}
D^2 = \sum^{n}_{i=1} \frac{[\sedobs(m_i) - \sedapp(m_i)]^2}{\sigma_{m_i}^2 + \sigma_{\mathrm{cal}}^2}
\label{d2}
\end{equation}
where $\sigma_{m_i}^2$ is the squared photometric uncertainty for each magnitude $m_i$ and $\sigma_{\mathrm{cal}}^2$ is an extra uncertainty term which we add to account for absolute calibration errors. This is required because the photometric uncertainty given by the IPHAS database only represents the relative uncertainty due to background noise, which is often smaller than 0.01\,mag for bright stars. In reality, ground-based survey photometry rarely reaches an absolute accuracy better than a few percent due to additional sources of noise (e.g. variable atmospheric conditions).

We found empirically that a value of $\sigma_{\mathrm{cal}}=0.1$ is required to prevent the expected match between the observed data and the modelled SED to be too exact. Leaving this term out has the effect of producing a complex multi-modal probability landscape which prevents the sampling algorithm --discussed below-- from converging efficiently. The term can be considered as a way to account for all those sources of noise which we could not explicitly model in the other parts of the model.

\subsection{Sampling the joint distribution using MCMC}
In the previous sections we defined the joint distribution $P(\param, \data)$ and thus the posterior $P(\param | \data)$ as a product of priors and likelihoods. At this point the question remains how this distribution is computed in practice. A simple ``brute force'' method which computes the model for the entire parameter space is computationally intractable. Even if we were to sample the parameter space at a coarse resolution, say, 100 values for each of the 10 model parameters, we would require the distribution to be computed in $10^{20}$ points. This would occupy the author's computer for many billion years.

Fortunately, probability distributions can be sampled efficiently using a specialised class of algorithms called \emph{Markov Chain Monte Carlo} (MCMC). In brief, MCMC algorithms generate a pseudo-random walk in the parameter space in such a way that, over time, points in the space are visited with a frequency that is proportional to a specified probability distribution. Only useful points in the parameter space tend to be evaluated and we avoid wasting time calculating an infinite number of points in the improbable regions.
The details of MCMC algorithms are beyond the scope of this paper but we recommend \citet{chib1995} for an introduction and \citet{mackay} and \citet{gregory2005} for background reading. 

Several libraries are available which allow Bayesian models to be defined and sampled using MCMC. In this work we used the {\sc PyMC} framework for Python \citep{pymc} which has the advantage that it allows models to be defined in a very concise way. Our annotated model takes less than two pages and is therefore included at the end of this paper for easy reference (Appendix~A). As such, this paper contains a precise and repeatable specification of the parameter estimation procedure. The source code and accompanying files are also available from the GitHub repository of the author\footnote{https://github.com/barentsen}.

For each object under study in this paper (to be explained in \S3), we sampled the joint distribution using the default settings of PyMC, which is to use the traditional \emph{Metropolis-Hastings} walking algorithm with a Gaussian \emph{step proposal function} (we refer to the manual of PyMC for definitions of these technical terms). PyMC automatically tunes the size of the step proposal function to ensure the walk is made efficiently with an \emph{acceptance rate} between 20 and 50\%.

It is usually sufficient to sample a probability distribution in only a few hundred independent points to obtain a sufficiently accurate approximation. MCMC algorithms typically require far more samples to be obtained however, because the walking algorithm naturally produces chains which are auto-correlated and hence may be stuck at local maxima. The true number of iterations which are required depends on the shape of the probability landscape; a complex or multi-modal distribution with sharp hills and valleys tends to require far more samples. In our application we find the chains to be auto-correlated over a typical length of 50 to 250 steps (depending on the properties of the star). For this reason, we decided to sample each object in 250\,000 points such that the total number of points is at least 3 orders of magnitude larger than the auto-correlation effect (i.e. at least 1000 truly independent samples are obtained for each object).

It is likely that our application would profit from recent advances in MCMC algorithms which claim to reduce the auto-correlation effect considerably. We draw the reader's attention to an implementation of such algorithm made available by \citet{foreman2012} which we intend to evaluate in future work.

\subsection{Results of the sampling procedure}
\label{sec:samplingresults}
\begin{figure*}
\includegraphics[width=\linewidth]{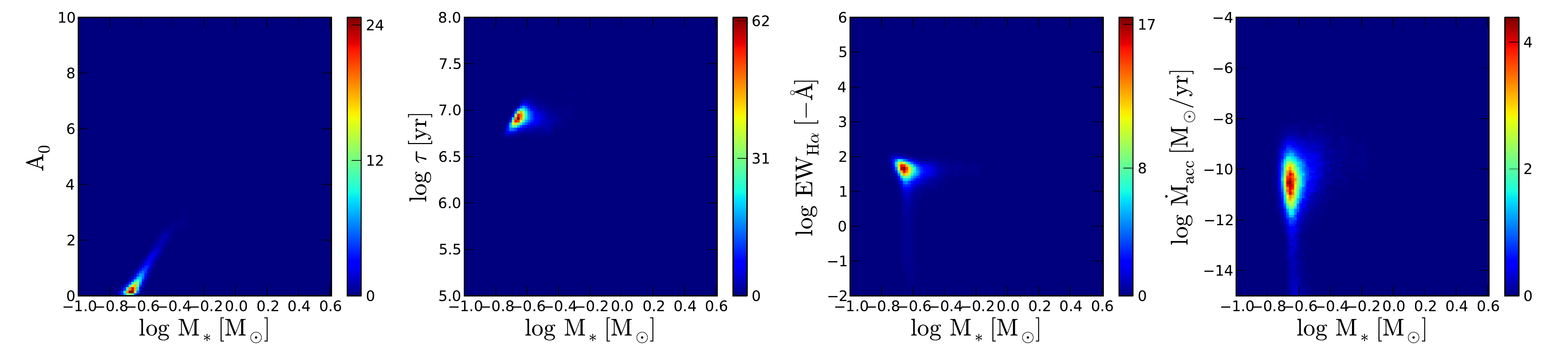}
\caption{Normalised 2D-histograms of the distribution of points in the MCMC chains for object C11059 (\mbox{\sloanr=19.4\,$\pm$\,0.03;} \mbox{\rha=1.3\,$\pm$\,0.04;} \mbox{\ri=1.9\,$\pm$\,0.04;} \mbox{\iJ=2.1\,$\pm$\,0.06}). These histograms trace the posterior distributions marginalised over all other parameters. Blue regions show areas in the parameter space with low probability, red areas show areas with high probability.
}
\label{example1}
\includegraphics[width=\linewidth]{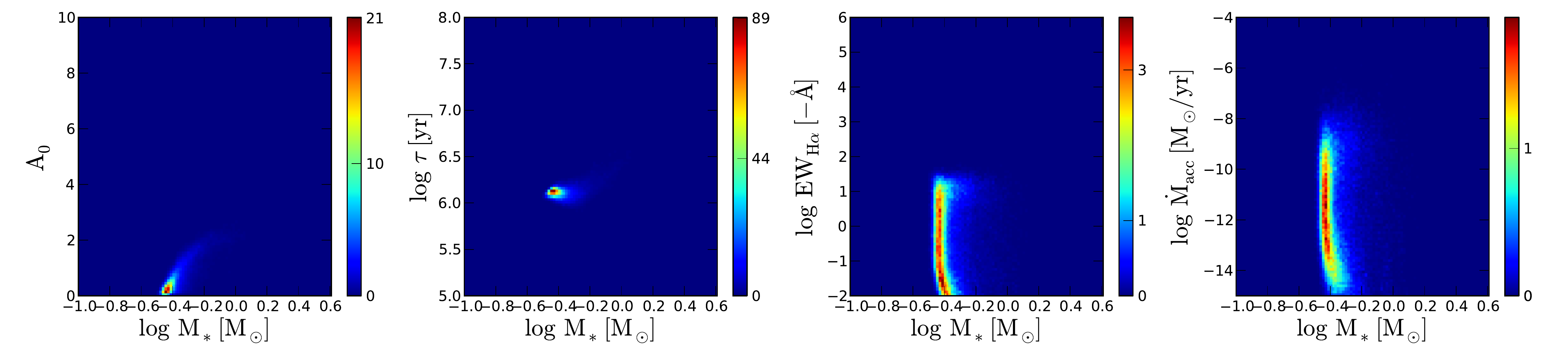}
\caption{Histograms for object C15152 (\mbox{\sloanr=16.3\,$\pm$\,0.002;} \mbox{\rha=0.8\,$\pm$\,0.004;} \mbox{\ri=1.4\,$\pm$\,0.003;} \mbox{\iJ=1.7\,$\pm$\,0.02}). Compared to the first example, this is a brighter object with near-zero \halpha\ emission and a slightly higher mass. The probability region is marginally more condensed owing to the smaller photometric uncertainties.}
\label{example2}
\includegraphics[width=\linewidth]{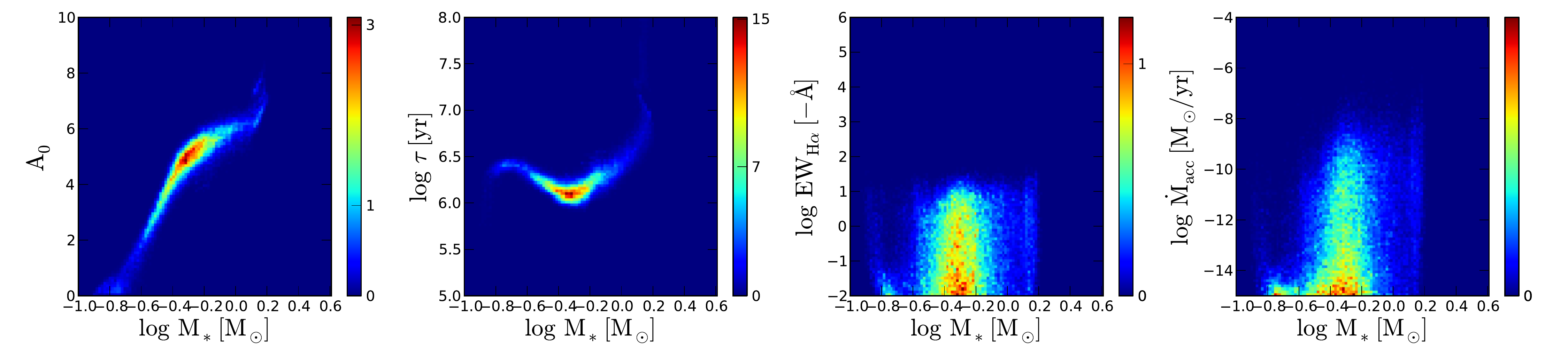}
\caption{Histograms for object C29493 (\mbox{\sloanr=19.3\,$\pm$\,0.04;} \mbox{\rha=0.7\,$\pm$\,0.08;} \mbox{\ri=2.0\,$\pm$\,0.05;} \mbox{\iJ=3.1\,$\pm$\,0.04}). Compared to the first example shown in Fig.~\ref{example1}, the \iJ\ colour is significantly redder which results in a higher extinction. We also draw attention to the banana-like shape of the high-probability region, which is a result of changes in the direction of the model isochrones relative to the reddening vector.
}
\label{example3}
\includegraphics[width=\linewidth]{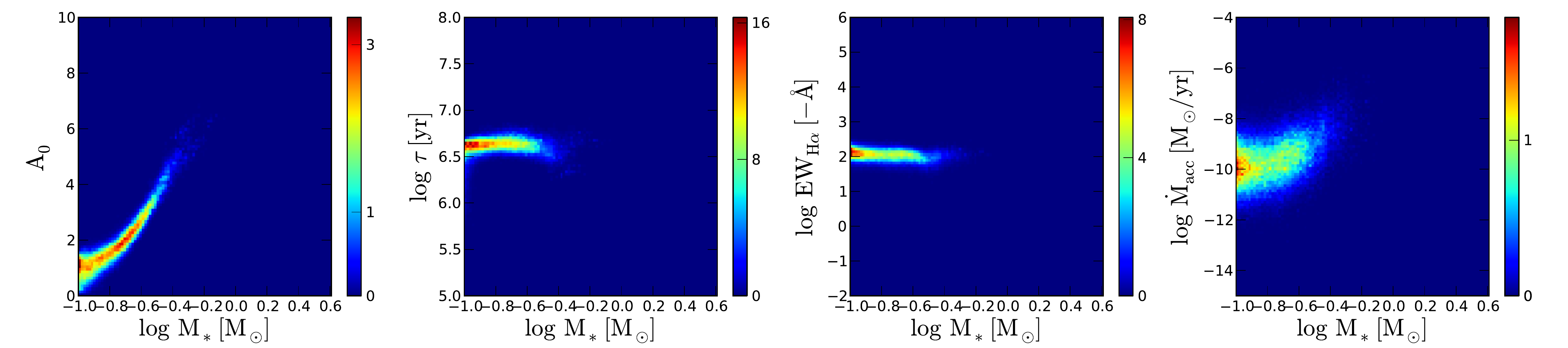}
\caption{Histograms for object C36902 (\mbox{\sloanr=20.6\,$\pm$\,0.09;} \mbox{\rha=1.6\,$\pm$\,0.11;} \mbox{\ri=2.1\,$\pm$\,0.10;} \mbox{\iJ=3.1\,$\pm$\,0.07}). This is one of the lowest-mass objects in our study. The high-probability region appears to go slightly beyond the range of the \citeauthor{siess} model.
}
\label{example4}
\end{figure*}
To verify the reliability of our procedure, the sampling was carried out using 5 independent walks of 50\,000 steps, each starting at randomised positions (with a \emph{burn-in length} of 1\,000 steps). We consistently found these independent chains to converge to the same parameter-space regions of high likelihood within a few hundred iterations, i.e. fast convergence towards a global maximum was reached in all cases.

We visualised the samplings by means of 2D-histograms (Figs.~\ref{example1}-\ref{example4}), which trace the posterior marginalised over all other parameters. The first two examples (Figs.~\ref{example1}-\ref{example2}) are representative for the vast majority of objects in our study which show low levels of extinction (\mbox{$\a0 < 1$}). The main difference is that the first example shows evidence for \halpha\ emission, whereas the second example does not.

Non-typical examples are shown in Figs.~\ref{example3}-\ref{example4}. These objects have (\iJ) colours which are significantly redder compared to the first two examples, in a way that is consistent with a higher level of extinction. We note that the uncertainties are significantly larger for these more highly reddened objects. This is a result of our decision to adopt a log-normal extinction prior with a peak near \a0=0.5 and a long tail towards higher values (corresponding to the empirical distribution for the region which we adopted in \S2.3).

If we had not included this prior information then the uncertainty in Figs.~\ref{example1}-\ref{example2} would have been similar to that in Figs.~\ref{example3}-\ref{example4}. In other words, whilst our dataset offers a rough constraint on the individual extinction, the degeneracy between extinction and mass is not resolved entirely and the prior makes a contribution towards constraining the result. At the same time, Figs.~\ref{example3}-\ref{example4} demonstrate that the prior does not prevent higher levels of extinction and uncertainty to be revealed when the data are inconsistent with low extinction.
We will quantify the contribution of the prior in the discussion at the end of the paper (\S\ref{sec:extinctionprior}).

We draw the reader's attention to the 'banana'-shaped posterior which appears when the uncertainty is large. This is a natural result of curves in the model evolutionary tracks relative to the reddening vector.

We also note that the range of possible masses in Fig.~\ref{example4} appears to go somewhat beyond the mass range that is provided by the \citeauthor{siess} model. Only a few faint objects are affected in this way however, and we decided not to remove these from our study.

Finally, we note that a visual inspection of all objects under study revealed that these marginalised posteriors distributions are single-moded and quasi-symmetric when the logarithm of each parameter is considered. This implies that the marginalised posterior can be well-characterised by means of computing expectation values and standard deviations. 

The posterior was sampled and characterised in this way for a total of 587 known members of the NGC\,2264 star-forming region. In the following sections, we discuss how the input dataset was obtained (\S3) and how the derived parameters make sense of the objects' positions in colour/magnitude diagrams (\S4-\S5).

\section{Application to NGC\,2264}
NGC\,2264, also known as the \emph{Cone Nebula} or \emph{Christmas Tree Cluster}, is located at a distance of $\sim$760\,pc in the constellation of Monoceros \citep{sung1997}. The cluster is estimated to contain $\sim$1000 members, most of which have been identified using a range of methods including \halpha\ and variability surveys, X-ray observations and mid-infrared imaging \citep[a review is given by][]{handbook2264}.

Stellar parameters have previously been estimated for members of the region using a wide variety of colour/magnitude diagrams and evolutionary models  \citep[e.g.][]{park2000,rebull2002,sung2004,flaccomio2006,dahm2007}. 
There are significant systematic differences between these studies however, with median age estimates for the cluster ranging between $\sim$1 and 5~Myr \citep{handbook2264}. 

Moreover, there is only a partial overlap in the membership samples considered in these previous studies, owing to differences in the datasets and member selection criteria. In what follows we attempt to select the largest and most homogeneous membership sample of NGC\,2264 to date, and then use it to demonstrate our method. 

\subsection{Membership list}
\label{membership}
The most comprehensive catalogue of objects towards NGC\,2264 has been presented by~\citet{sung2008,sung2009} hereafter referred to as S08 and S09. Their work is based on a compilation of 
\begin{enumerate}
\item deep \textit{VRI} and \halpha\ photometry using the 3.6\,m Canada France Hawaii Telescope (CFHT); 
\item bright \textit{BVRI} and \halpha\ photometry using the 1\,m telescope at Siding Spring Observatory (SSO);
\item low- and moderate-resolution literature spectroscopy from \citet{reipurth2004} and \citet{dahm2005};
\item archival X-ray observations from the Chandra and ROSAT Space Telescopes;
and \item archival infrared observations from the Spitzer Space Telescope.
\end{enumerate}

The authors constructed a catalogue of 69\,674 optical objects detected towards the cluster (tables 3, 8 \& 9 in S08) and then assigned various ``membership codes'' by crossmatching the \halpha, X-ray and infrared data. We used these codes to select a total of 1191 likely members which satisfy one or more of the following criteria:
\begin{enumerate}
\item \halpha\ emission stronger than chromospherically active main-sequence stars (membership code: `H', `E', `+', `P' or `p');
\item strong X-ray emission (code: `X', `+', `-', `M' or `P');
\item Spitzer colours consistent with a protoplanetary disc (code: `I', `II', `II/III', `pre-TD' or `TD' in S09).
\end{enumerate}

In this sample of 1191 objects, 42\% satisfy the \halpha\ criterion, 62\% satisfy the X-ray criterion, and 38\% satisfy the Spitzer criterion. 
There is considerable overlap: 32\% satisfy more than one criterion while 10\% satisfy all three. 

Additional signatures of membership such as radial velocities, or chemical indicators of youth such as Lithium, are currently not available for most of these objects. This is, in part, because half of the sample is fainter than V$>$18 for which high-resolution spectroscopy becomes increasingly expensive. Based on the clustering properties and positions in the colour-magnitude diagram however, S08 convincingly argued that the vast majority of objects in this sample are genuine members of NGC\,2264. Nevertheless, it is likely that our sample contains a small number of foreground/background objects.

\begin{figure*}
\includegraphics[width=\linewidth]{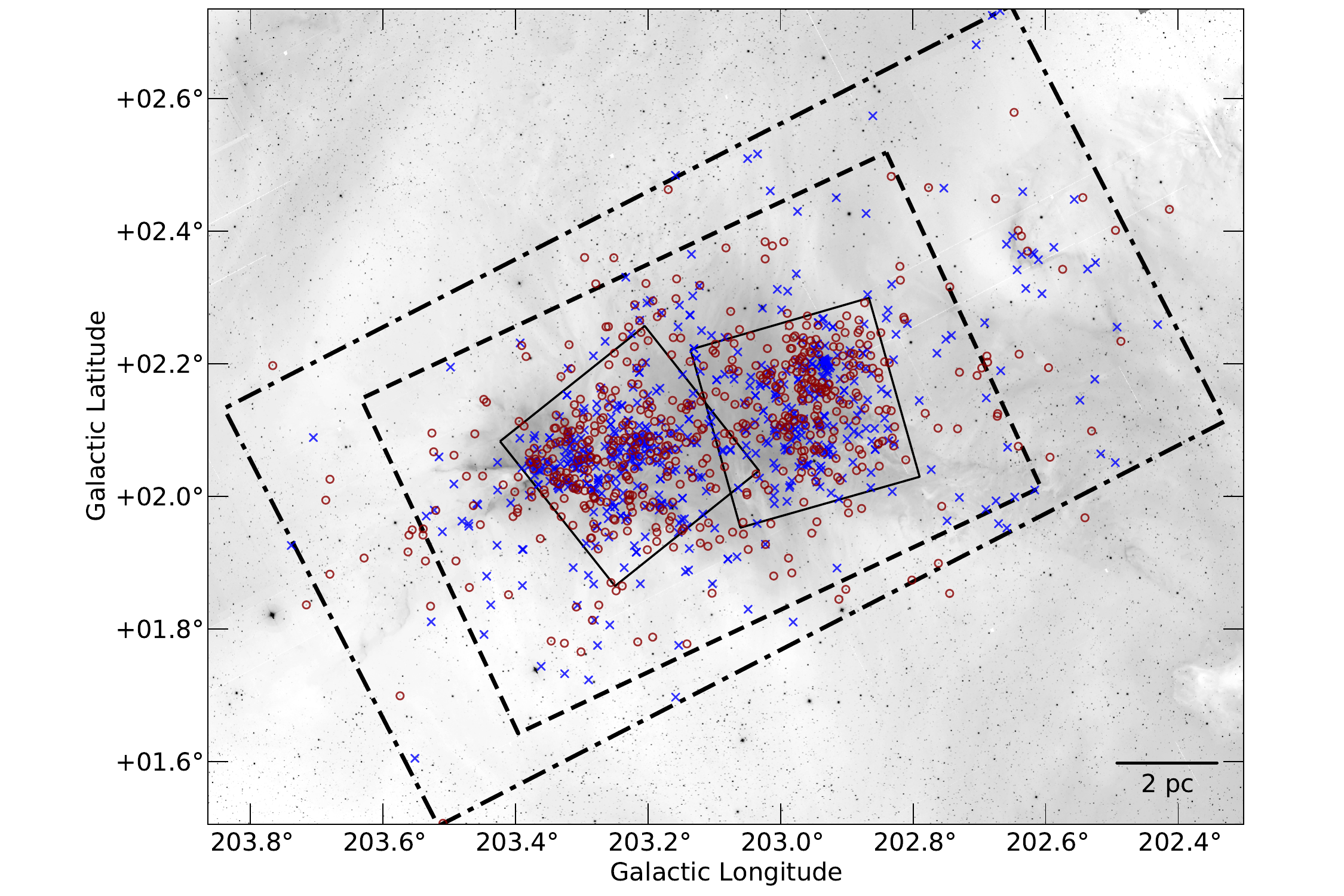}
\caption{Spatial distribution of known candidate members in NGC\,2264 taken from the works by Sung et al. (see text.) Red circles show objects for which high-reliability IPHAS and 2MASS photometry is available which satisfy the strict quality requirements defined in \S2.4; these are the objects which are studied in our work. Blue crosses show candidate members for which we could not obtain high-quality IPHAS/2MASS photometry. Large rectangles show the footprints of the deep optical CFHT observations (dash-dotted line), Spitzer IRAC observations (dashed line) and Chandra observations (solid line). A few objects fall outside these footprints, because they were included on the basis of bright optical photometry or literature spectroscopy (footprints not shown). The background image is an inverted version of Fig.~\ref{fig:map}.}
\label{fig:map_members} 
\end{figure*}

The spatial distribution of the sample is shown in Fig.~\ref{fig:map_members} together with the footprints of the observations from which the sample was compiled.
\begin{figure*}
\includegraphics[width=\linewidth]{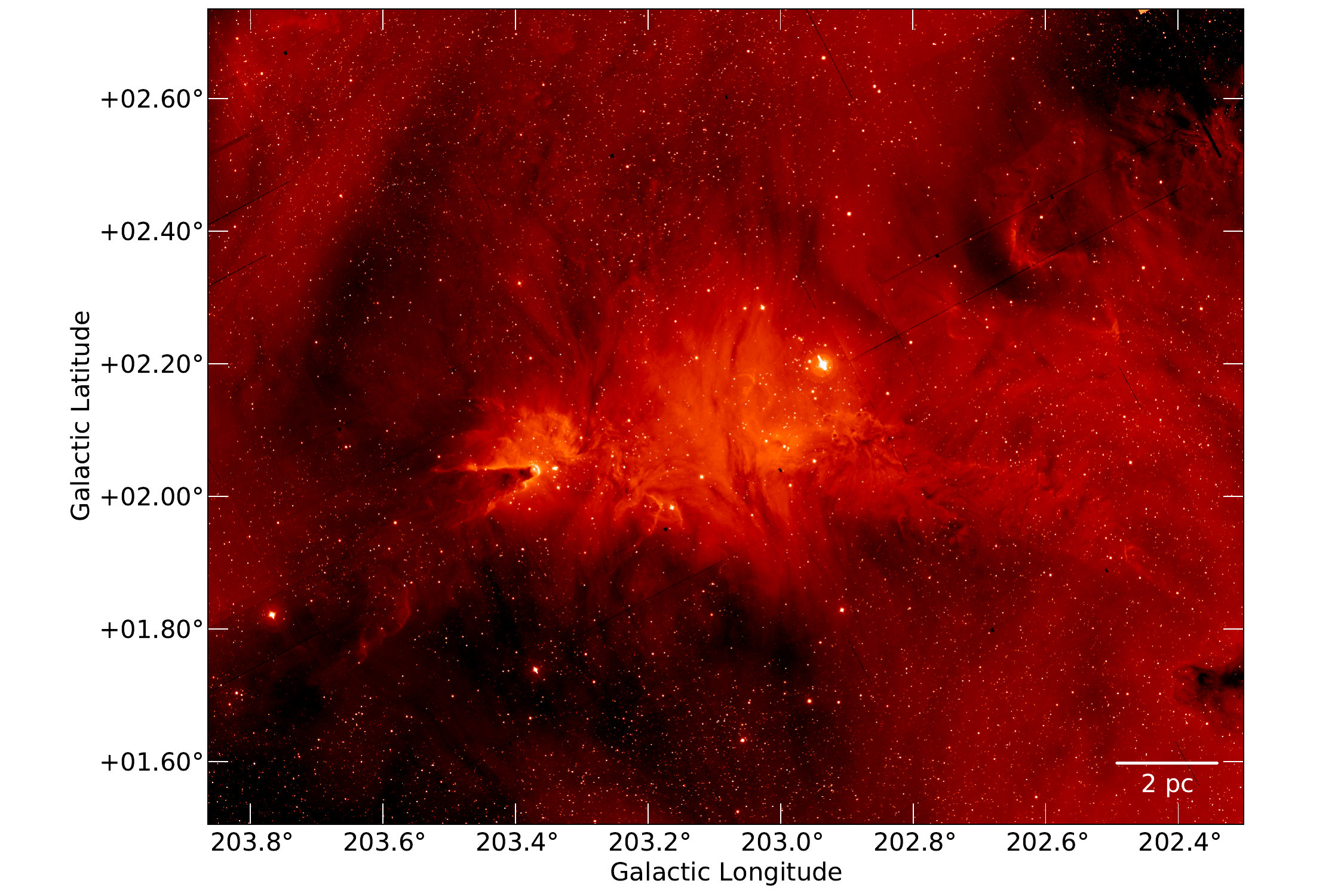}
\caption{Mosaic of IPHAS observations towards NGC\,2264 in the \halpha\ band. A few narrow black bands can be seen, which indicate small gaps of missing data which will be completed in a future IPHAS data release.}
\label{fig:map} 
\end{figure*}

\subsection{IPHAS counterparts}
IPHAS is a 1800 \sqdeg\ photometric survey of the Northern Galactic Plane 
($30^{\rm o} \lesssim \ell \lesssim 220^{\rm o}$, $-5^{\rm o} \lesssim b \lesssim +5^{\rm o}$) carried out using a narrow-band \halpha\ filter and the broad-band Sloan \sloanr\ and \sloani\ filters, using the 0.3\,\sqdeg\ Wide-Field Camera (WFC) at the 2.5-meter Isaac Newton Telescope (INT) in La Palma. Data towards NGC\,2264 were obtained as part of the survey during several observing runs between 2003 and 2009. The central part of the cluster, containing the vast majority of members, was observed on 2008 January 17 with an average seeing of $1.2\pm0.1$\arcsec (IPHAS field numbers 3773 and 3773o).

We used the {\sc Montage} toolkit to create an \halpha\ mosaic of 20 fields towards the region which is shown in Fig.~\ref{fig:map}. The contrast of the mosaic has been stretched using an arcsinh curve to bring out the \halpha\ background emission.

All data were pipeline processed at the Cambridge Astronomical Survey Unit (CASU) as detailed in \citet{irwin2001}, \citet{drew2005} and \citet{idr}. This routinely includes photometric calibration based on nightly standard star fields, with all magnitudes based on the Vega system. In addition, we have been able to draw upon the results of a global calibration of the survey data which significantly reduces field-to-field magnitude shifts (Drew et al, in preparation).

We crossmatched the sample of 1191 members from S08 against the IPHAS catalogue.
The astrometry of both S08 and IPHAS is based on 2MASS reference coordinates which offers a typical accuracy of 0.1 to 0.2\arcsec\ \citep{idr}.
For this reason, we decided on a strict matching distance upper bound of 0.5\arcsec. A total of 819 members were found to have a counterpart within the matching distance in \emph{all} three IPHAS bands. For 72 of these objects \halpha\ photometry was not previously available in the S08 catalogue.

From the 372 objects which could not be matched in all three bands, 249 fall below the typical detection limit of IPHAS ($R > 20$) and 44 are saturated ($R < 13$). Most of the remaining objects were found to be blended with a nearby neighbour in IPHAS while being resolved in the higher resolution CFHT-based data from S08, which produces an astrometric offset.
Increasing our matching distance to 1.0\arcsec\ would include 35 of these objects, but we decide against this in favour of data reliability.

In the majority of cases the matched objects were detected 2 or 3 multiple times by the IPHAS survey, usually in the same night. This is because IPHAS fields are observed twice with a small offset to account for gaps between CCD chips in the camera. When two or more detections of the same object are available, we derived the mean magnitude and updated the uncertainty.

\subsection{New candidate members from IPHAS}
Having obtained a sample of 819 very likely members from literature, we carried out a search in the IPHAS database for any new \halpha\ emitters which may have been missed during previous searches. Using the method detailed in Paper~I we identified a total of 164 objects which are located confidently above the main locus of stars in the IPHAS (\ri)/(\rha) diagram. The selection threshold explained in Paper I is designed to avoid chromospherically active field stars, and so the vast majority of these objects are likely to be accreting T Tauri stars.

We find that 150 (91\%) of these objects were already selected because they are classified as strong \halpha\ emitters in S08, while a further 9 objects were selected on the basis of their X-ray or infrared emission but not on the basis of \halpha\ (object IDs: C27213, C34019, C36055, C37804, C39811, C42487, C8431, W3407, W5604). Only 5 objects were not already selected and are added to our sample (C18650, C19751, C25302, W2320, W5620). This brings the sample size to 824 objects. The fact that we are unable to identify a significant number of new \halpha-emitters suggests that the sample is very complete in this respect, at least down to the detection limits of IPHAS.

\subsection{Quality criteria}
\label{quality}

At this point we could continue our investigation with all of the 824 
objects for which IPHAS counterparts were found. However, we choose to narrow down the sample to 587 objects using the following strict quality requirements:
\begin{enumerate}
\item the photometric uncertainty on each of the three IPHAS magnitudes must be smaller than 0.1;
\item each object must be classified as ``strictly stellar'' or ``probable stellar'' in the IPHAS catalogue in all three bands \citep[quality flag `-1' or `-2' defined by][]{idr};
\item the object must have a J-band counterpart in the 2MASS database with Signal-to-Noise Ratio (SNR) $>$ 5 (quality flag `A', `B' or `C');
\item the object must not be marked as an unresolved binary in the catalogue due to S08 or 2MASS (flag `D') and its nearest neighbour must be further away than 1 arcsecond.
\end{enumerate}

Criterion (i) avoids faint sources with high uncertainties. The criterion corresponds to a SNR of 10 or typical magnitude limits of 20.5 in \sloanr\ and 19.5 in \sloani/\halpha\ \citep[][]{idr}, although we note that there are small spatial variations in the true magnitude limits depending on the observing conditions and the number of repeat observations. Only 34 objects did not meet this criterion.

Criterion (ii) deals with the issue of flux-contamination. The magnitudes in the IPHAS database are based on aperture photometry which is prone to contamination by nearby neighbours or spatially varying background emission. The IPHAS pipeline solves this problem by tracking variations in the background emission on scales of 20-30 arcsec. In some cases the background changes on scales smaller than 5-10 arcsec however, in which case photometric measurements of faint sources become unreliable.

Fortunately, photometry which is unreliable for this reason is flagged in the pipeline as part of the morphological classification step (see \S2.1 in Paper~I). In brief, the pipeline derives a curve-of-growth for each object from a series of aperture flux measures with different aperture radii. When this curve deviates from the characteristic point spread function (PSF) of other stars in the field, the object is flagged as ``extended'' (class 1) or ``probable extended'' (class -3).

By requiring objects to be classified as ``strictly stellar'' (code -1) or ``probable stellar'' (code -2) in all three bands, we ensure that only reliable measurements for objects with a normal-shaped PSF are included in our analysis. A total of 143 objects did not meet this criterion.

Criterion (iii) is introduced because IPHAS magnitudes alone are not sufficient to constrain the extinction of individual objects as we explained in \S2. A total of 62 objects did not meet this criterion. Future work could benefit from the deeper J-band data offered by the UK Infrared Deep Sky Survey \citep[UKIDSS,][]{lucas2008}, but data for NGC\,2264 is not yet available from that survey at this time.

Finally, criterion (iv) avoids likely problems due to source confusion. A total of 30 objects did not meet this criterion.

After applying each of the criteria, we are left with 587 objects (because a number of objects failed more than one criterion). The resulting table of IPHAS and 2MASS photometry for these objects is given in Table~\ref{tab:phot} and forms the basis for our analysis.

\section{Results}

We obtained Bayesian posterior distributions for each of the 587 objects selected above. We then summarised the posteriors by computing marginalised means and standard deviations, i.e. we derived a point estimate per parameter for each object. These values are listed in Table~\ref{tab:param} and visualised by histograms in Fig.~\ref{fig:hist}. 
In this section we provide a brief overview of the results by inspecting the distributions of these point estimates. 
This will lead us to a set of preliminary results on the properties of NGC\,2264, such as its mass distribution, median age and fraction of accretors.

Whilst point estimates of stellar parameters are widely used as a tool to investigate the properties of a cluster, we warn that the presence of large uncertainties can make a direct analysis of point estimates unreliable. 
For example, while it is tempting to estimate the age of NGC\,2264 from the histogram of mean stellar ages shown in Fig.~\ref{fig:hist}, the result may be biased due to the presence of assymetric and correlated uncertainties  \citep[see][]{pont2004}. 
This does not imply that our results cannot be used to study the properties of the cluster, in fact our posteriors contain all the required information on the uncertainties and degeneracies. In order to exploit this information however, we would need to build a probabilistic model which links the global parameters of the cluster to the posteriors of the stars.

Whilst it is easy to see that our hierarchical model may be extended to include the global properties of the cluster as parameters, the priors and likelihoods of such parameters would have to be chosen with care. 
They would need to reflect the current knowledge in the field and allow the right questions to be answered. 
For example, if we were to estimate the age and age spread of the cluster, we would need to make a detailed appraisal of the accuracy of pre-main sequence isochrones and include the information in the model.

Defining such a cluster model is beyond the scope of the present work. In what follows we merely offer the reader a concise summary of the distribution of the point estimates, which may be considered as a first-order approximation of the global properties of NGC\,2264. In \S6 we will discuss the future prospect of extending our work to obtain a complete model of the cluster.

\label{results}

\subsection{Masses, ages \& extinction}
\begin{figure*}
\includegraphics[width=\linewidth]{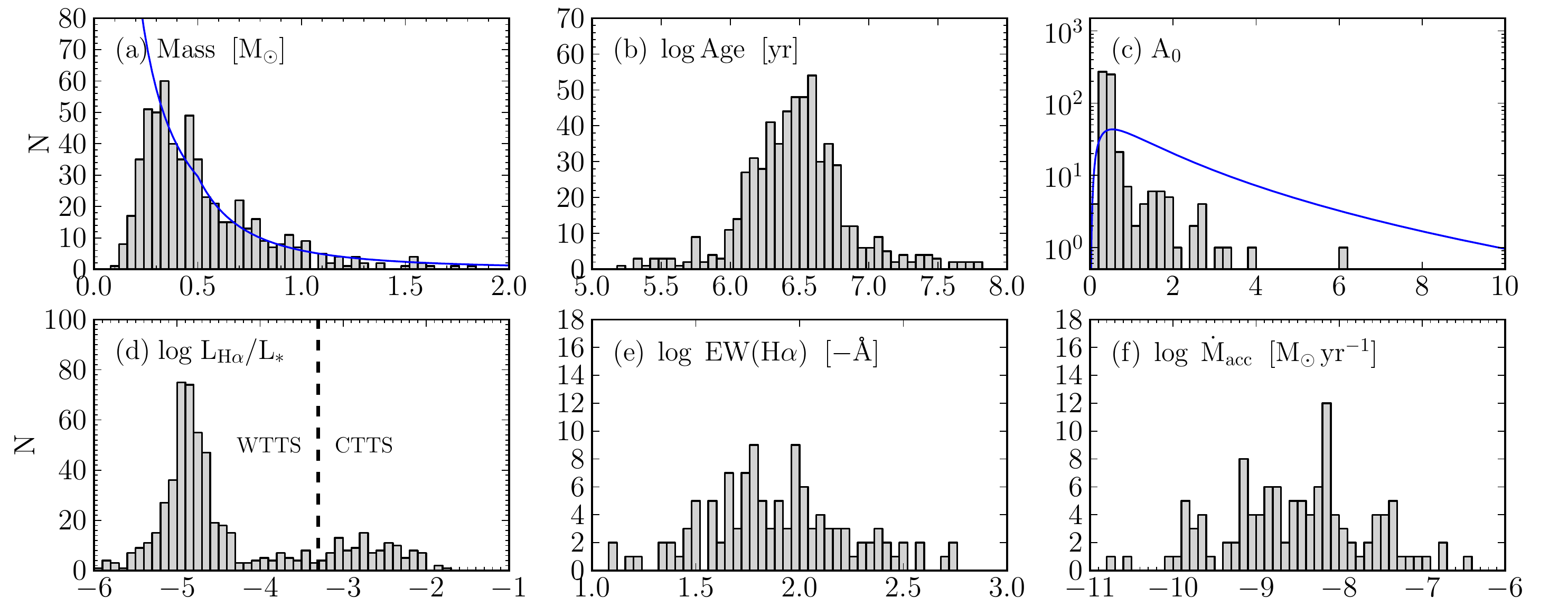}
\caption{Distribution of inferred parameters listed in Table~\ref{tab:param}. Blue solid lines in panels (a) and (c) show the priors. We note that panels (e) and (f) only include the CTTS objects. We warn that these histograms do not reflect the underlying uncertainties and degeneracies.}
\label{fig:hist} 
\end{figure*}

The mass distribution (Fig.~\ref{fig:hist}, panel a) shows the expected power-law increase towards lower masses. Compared with the Kroupa IMF (blue solid line) we find a deficit of objects with masses below 0.25\,\msol. Our sample is significantly less complete below this mass due to the SNR quality criteria imposed in \S2.4. Similarly, many stars with masses heavier than $\sim$1.2~\msol\ are missing due to the saturation limits of the IPHAS survey. The highest inferred mass in our sample equals $1.8^{+0.3}_{-0.2}$\,\msol.

The age distribution (panel b) shows a mean age of $\log\tau=6.48 \pm 0.38$ (which corresponds to the median $\tau = 3.0$\,Myr), albeit with a large apparent dispersion between 1.8 and 4.5 Myr (25 and 75\% quartiles). Our median estimate of 3~Myr is identical to the main-sequence turn-off age obtained in the seminal paper by \citet{walker1956}, and is also consistent with the age of 3.1~Myr reported by \citet{sung2004} using the same set of isochrones as in our work.

The extinction (panel c) shows a mean of $\log A_0=-0.37\pm0.20$ ($A_0=0.43$) which is consistent with the widely reported low levels of foreground extinction. 30 objects show higher levels of extinction ranging between \a0=1 and 3, while only 4 objects show extinctions beyond $\a0>3$ (beware that the vertical axis in panel c is logarithmic for clarity).

Compared to the log-normal extinction prior (solid blue line) there is a  deficit of objects with large extinctions, which will be discussed in \S\ref{sec:extinctionprior}.

\begin{table*}
\begin{tabular}{llccccccc}
\hline
  Name & IPHAS ID         & \multicolumn{3}{c}{IPHAS magnitudes} & \multicolumn{3}{c}{2MASS magnitudes} \\ 
  (S08) & J[RA(2000)$+$Dec(2000)] & \sloanr & \halpha & \sloani & J & H & K \\
\hline  
C11059 & J063955.90+094239.8 &  19.41$\pm$0.03 & 18.14$\pm$0.03 & 17.56$\pm$0.02  & 15.44$\pm$0.06 & 14.69$\pm$0.05 & 14.28$\pm$0.08 \\
C11997 & J063957.95+094104.7 &  17.91$\pm$0.01 & 16.50$\pm$0.01 & 16.16$\pm$0.01  & 14.36$\pm$0.04 & 13.65$\pm$0.03 & 13.35$\pm$0.04 \\
C12135 & J063958.30+092848.6 &  18.51$\pm$0.01 & 17.66$\pm$0.02 & 16.74$\pm$0.01  & 14.99$\pm$0.04 & 14.34$\pm$0.05 & 14.11$\pm$0.07 \\
C12598 & J063959.23+100607.7 &  17.26$\pm$0.00 & 16.18$\pm$0.01 & 16.07$\pm$0.00  & 13.91$\pm$0.02 & 12.80$\pm$0.03 & 12.10$\pm$0.02 \\
C13507 & J064001.30+094300.5 &  19.10$\pm$0.04 & 16.91$\pm$0.01 & 18.14$\pm$0.04  & 15.00$\pm$0.04 & 13.17$\pm$0.02 & 11.94$\pm$0.03 \\
C14005 & J064002.67+093524.3 &  16.86$\pm$0.00 & 15.87$\pm$0.00 & 15.35$\pm$0.00  & 13.70$\pm$0.03 & 12.99$\pm$0.03 & 12.71$\pm$0.03 \\
C15152 & J064005.22+095056.6 &  16.27$\pm$0.00 & 15.50$\pm$0.00 & 14.87$\pm$0.00  & 13.19$\pm$0.02 & 12.48$\pm$0.02 & 12.26$\pm$0.02 \\
C15247 & J064005.53+092226.1 &  16.54$\pm$0.00 & 15.65$\pm$0.00 & 15.44$\pm$0.00  & 13.95$\pm$0.03 & 13.12$\pm$0.02 & 12.73$\pm$0.03 \\
C15285 & J064005.53+094554.8 &  18.54$\pm$0.02 & 18.14$\pm$0.04 & 17.16$\pm$0.02  & 15.14$\pm$0.05 & 14.28$\pm$0.04 & 13.93$\pm$0.05 \\
C15519 & J064006.00+094942.7 &  16.44$\pm$0.00 & 15.62$\pm$0.00 & 15.18$\pm$0.00  & 13.46$\pm$0.03 & 12.76$\pm$0.03 & 12.45$\pm$0.03 \\

$\cdots$  & & & & & & & \\
\hline 
\end{tabular}
\caption{IPHAS and 2MASS photometry for known members of NGC\,2264 which satisfy our selection and quality criteria (see text).
The first column shows the existing object identifier as defined by \citet{sung2008}, while the second column shows the IAU-registered naming convention for objects detected by the IPHAS survey, which is formed by prefixing ``IPHAS'' to the position string.
Calibrated IPHAS photometry is given in columns 3-5
and matched 2MASS data is given in columns 6-8.
This table is available in its entirety in the online journal.}
\label{tab:phot}
\begin{tabular}{lccccccl}
\hline
 Name & $\log\a0$   & $\log \mass$ & $\log \tau$ & $\log \ewha$ &  $\log \lha$ & $\log \macc$ & Comments \\ 
 & [mag] & [\msol] & [yr] & [-\AA] & [\lsol] & [$\rm M_\odot\,yr^{-1}$] &  \\
\hline
C11059 & $-0.41 \pm 0.44$ & $-0.61 \pm 0.08$ & $6.91 \pm 0.09$ & $1.45 \pm 0.54$ & $-4.5 \pm 0.6$ &  & \\
C11997 & $-0.39 \pm 0.40$ & $-0.49 \pm 0.07$ & $6.50 \pm 0.10$ & $1.79 \pm 0.16$ & $-3.6 \pm 0.2$ & $-9.1 \pm 0.8$ & CTTS\\
C12135 & $-0.33 \pm 0.38$ & $-0.51 \pm 0.07$ & $6.77 \pm 0.09$ & $-0.20 \pm 0.85$ & $-5.8 \pm 0.9$ &  & \\
C12598 & $0.27 \pm 0.39$ & $-0.18 \pm 0.17$ & $6.56 \pm 0.31$ & $1.59 \pm 0.86$ & $-3.1 \pm 1.0$ & $-8.5 \pm 1.4$ & CTTS\\
C13507 & $0.19 \pm 0.73$ & $-0.34 \pm 0.28$ & $6.83 \pm 0.44$ & $1.41 \pm 2.86$ & $-3.8 \pm 3.0$ & $-8.8 \pm 3.3$ & CTTS\\
C14005 & $-0.42 \pm 0.42$ & $-0.41 \pm 0.07$ & $6.28 \pm 0.07$ & $0.49 \pm 1.13$ & $-4.5 \pm 1.2$ &  & \\
C15152 & $-0.33 \pm 0.40$ & $-0.36 \pm 0.09$ & $6.12 \pm 0.08$ & $-0.61 \pm 1.18$ & $-5.3 \pm 1.2$ &  & \\
C15247 & $-0.48 \pm 0.41$ & $-0.29 \pm 0.10$ & $6.59 \pm 0.19$ & $1.00 \pm 0.98$ & $-3.9 \pm 1.0$ &  & \\
C15285 & $0.21 \pm 0.27$ & $-0.32 \pm 0.14$ & $7.04 \pm 0.24$ & $-0.86 \pm 0.92$ & $-6.2 \pm 0.9$ &  & \\
C15519 & $-0.36 \pm 0.45$ & $-0.32 \pm 0.12$ & $6.27 \pm 0.13$ & $-0.44 \pm 1.44$ & $-5.2 \pm 1.5$ &  & \\

$\cdots$  & & & & & & & \\
\hline 
\end{tabular}
\caption{Posterior expectation values and standard deviations for parameters of NGC\,2264 members, obtained from IPHAS and 2MASS photometry using Bayesian inference. Shown here are the first 10 entries,
the table is available in its entirety in the online journal.}
\label{tab:param}
\end{table*}

\subsection{\halpha\ emission \& accretion rates}

The \halpha\ luminosities (panel d) are shown as a fraction of the object's bolometric luminosity $\mathrm{L_*}$ which we derived from the \citeauthor{siess} model as a function of an object's age and mass. This allows us to show the distribution relative to the \emph{chromospheric saturation limit} at $\mathrm{\log L_{H\alpha}/L_* = -3.3}$ (dashed line). This is the maximum level of \halpha\  emission observed in clusters at the age of 65 to 125~Myr, where accretion is thought to have ceased and \halpha-emission is produced entirely by chromospheric activity \citep{barrado2003}. 

This saturation limit is a widely used criterion to separate accreting from non-accreting objects. In the remainder of this paper we refer to objects which fall above the limit as ``accretors'' or \emph{Classical T Tauri Stars} (CTTS) while those which fall below the limit are ``non-accretors'' or \emph{Weak-lined T Tauri Stars} (WTTS). 

According to this definition, the accretion fraction is $20\pm2\%$ (115 objects). This fraction may be slightly underestimated because 36 additional objects fall only just below the criterion ($\mathrm{\log L_{H\alpha}/L_*}$ between -4.0 and -3.3). It is likely that some of these objects are undergoing very low levels of mass accretion which we cannot distinguish from chromospheric activity using our dataset. If we were to assume that all these objects are undergoing accretion then the fraction of accreting stars would rise to 25\%. We note that a frequency of 20 to 25\% for a cluster of 3~Myr is in excellent agreement with other clusters of a similar age \citep{fedele2010}.

The \halpha\ EW distribution for the accreting stars (panel e) shows a range from -12 to -546\,\AA. The corresponding mass accretion rates (panel f) range from $10^{-10.7}$ to $10^{-6.4}$\,\msol/yr with a median of $10^{-8.4}$\,\msol/yr. They are in broad agreement with accretion rates found in clusters of a similar age \citep[e.g.][]{sicilia2010}.

\section{Discussion}

The results presented in this paper can be used to investigate a wide range of questions regarding star formation and the history of NGC\,2264. However, in the remainder of this paper we choose to restrict ourselves to an evaluation of the method with an eye on future improvements. In what follows we show that the results obtained match (i) those expected from traditional colour-colour and colour-magnitude diagrams, and (ii) those previously reported in the literature using spectroscopy. We also discuss a small number of objects with anomalous colours and discuss the influence of the extinction prior.

\subsection{Comparison with colour/magnitude diagrams}

\begin{figure}
\includegraphics[width=\linewidth]{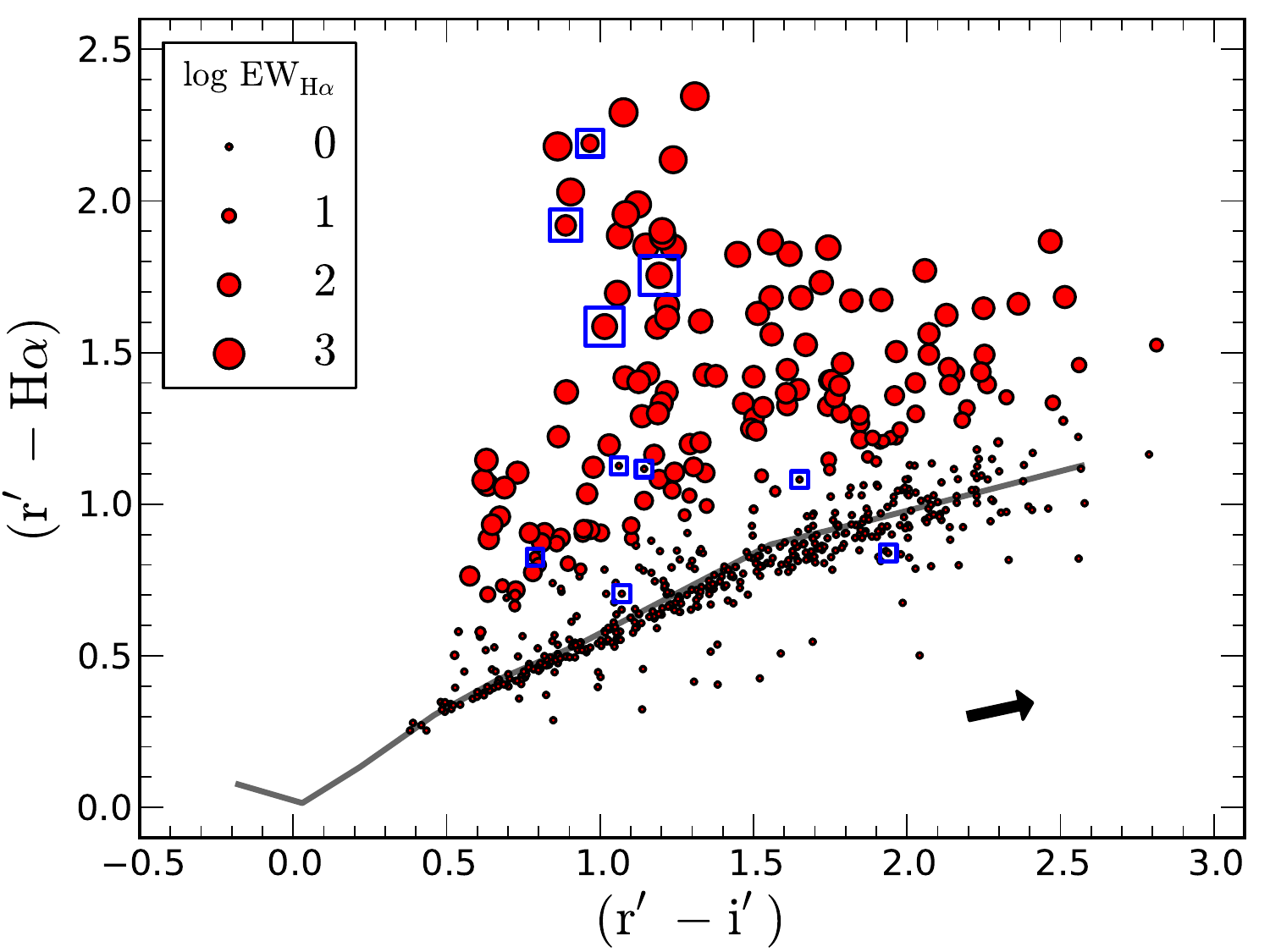}
\caption{Position of objects in the IPHAS (\ri)/(\rha)\ plane. The size of the circles represent the expectation value of the \halpha\ EW posterior. The solid line shows the unreddened main sequence while the arrow shows the typical reddening vector for a unit $\rm A_V$, both taken from Paper~I. Blue squares indicate objects with low likelihoods (cf. \S\ref{sec:lowlikelihood})}
\label{fig:valid_ccd} 
\end{figure}
\begin{figure}
\includegraphics[width=\linewidth]{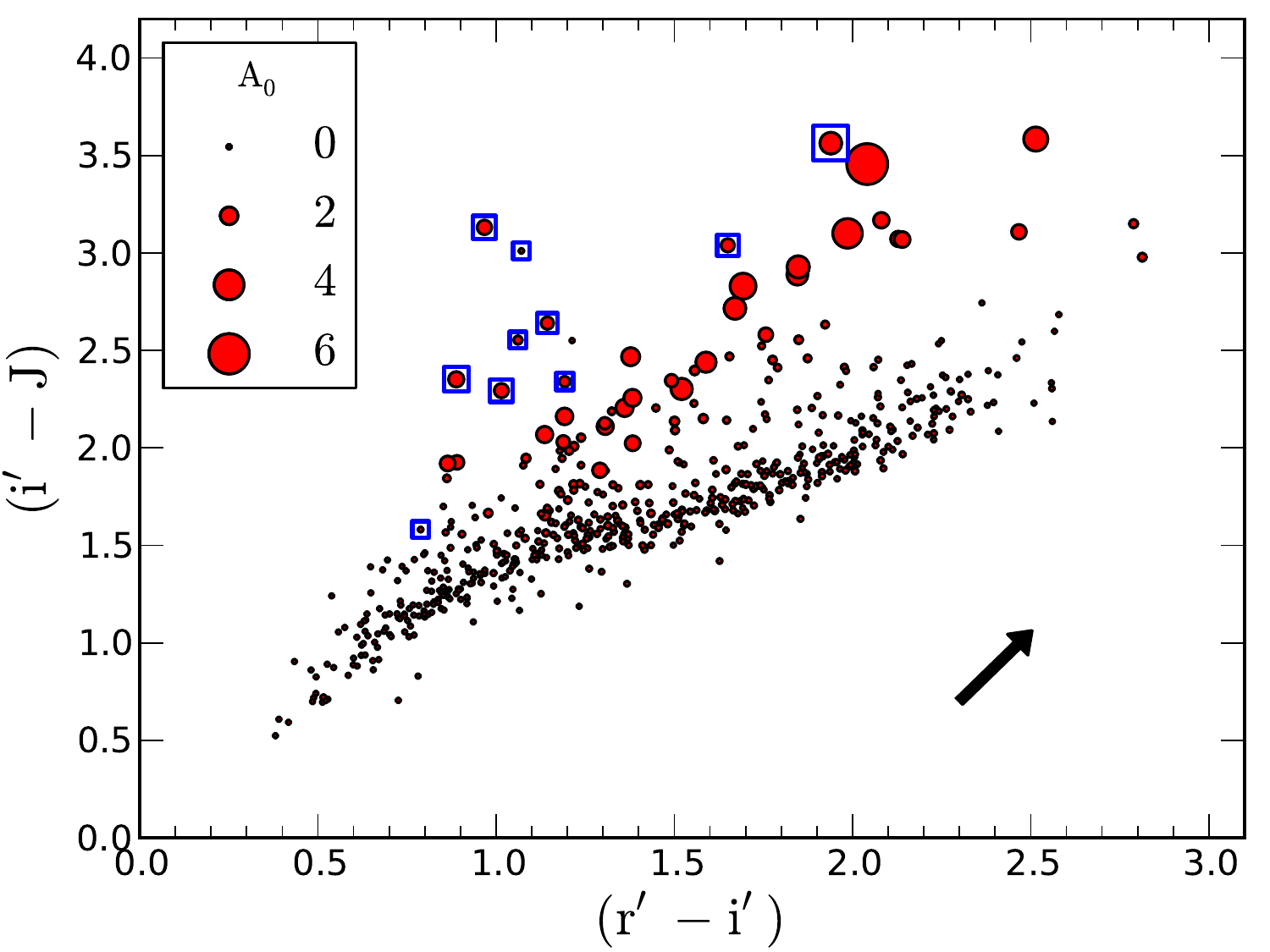}
\caption{Position of objects in the IPHAS-2MASS (\ri)/(\iJ)\ plane. The size of the circles represent the mean extinction posterior. The arrow shows the reddening vector for $\rm A_V=1$ due to \citet{schlegel}.}
\label{fig:valid_ccd_reddening}
\end{figure}

To verify that our results are consistent with those which would have been obtained using traditional plane-fitting methods, we show the position of objects as a function of their properties in three relevant colour/magnitude diagrams

First, Fig.~\ref{fig:valid_ccd} shows the (\ri)/(\rha)\ plane. This diagram acts mainly as an indicator for the \halpha-line strength: objects with \halpha\ in emission show greater \rha\ values and are therefore located above the main locus. The size of the symbols represent our posterior mean estimate for \ewha. We note the good correspondence between these estimates and the distance of an object from the main locus.

Second, Fig.~\ref{fig:valid_ccd_reddening} shows the (\ri)/(\iJ)\ plane. In this diagram we let the size of the symbols represent the extinction estimate, because the unit reddening vector (black arrow) follows a direction which is somewhat offset from the main locus of stars, such that objects with high extinction tend to be located above the main locus.

At first glance, this diagram shows a good agreement between the position of objects and their estimated extinction. However, we draw the reader's attention to the lack of a one-on-one relationship between the extinction and the apparent distance from the main locus. As discussed in \S\ref{sec:degeneracy}, this is a result of the fact that the \halpha-line falls inside the \sloanr-band, which has the effect of moving emission-line objects towards the left of the diagram and above the main locus, such that strong \halpha\ emitters with low extinction occupy the same region in the diagram as objects with weak emission and high extinction. This illustrates the fact that it is difficult to simultaneously estimate extinction and \halpha\ emission from these diagrams, which was a major motivation for adopting the Bayesian approach (cf. \S2).

\begin{figure}
\includegraphics[width=\linewidth]{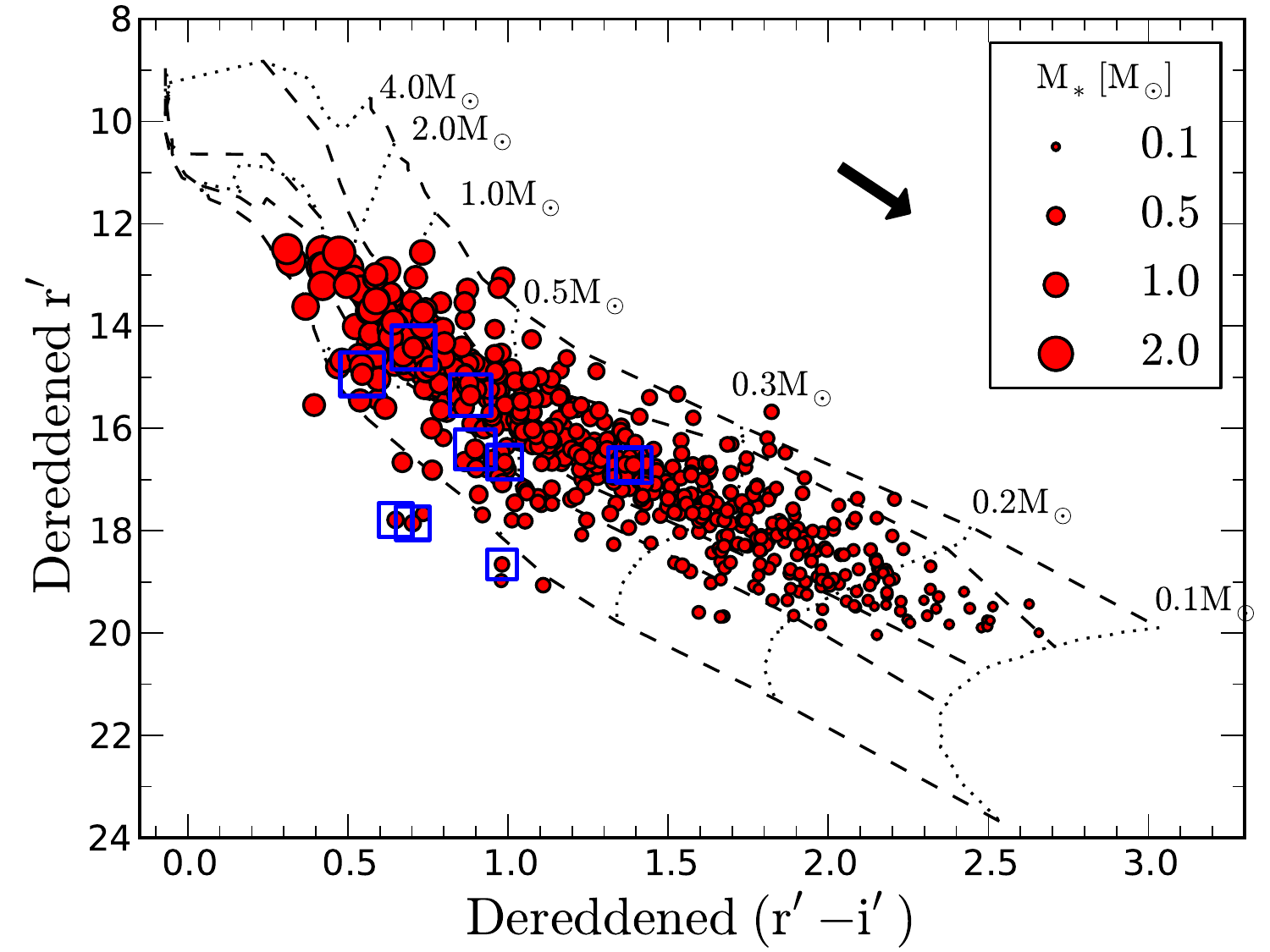}
\caption{Position of objects in the IPHAS (\ri)/\sloanr\ plane, dereddened using the mean extinction posterior values of each object. The size of the circles represent the mean stellar mass posterior. The arrow shows the reddening vector for a unit $\rm A_V$ due to \citet{schlegel}. We also show the evolutionary mass tracks (dotted lines) and isochrones (dashed lines) from the \citeauthor{siess} model, placed at the adopted distance of 760\,pc. The isochrones are for 0.1, 1, 5, 10 and 100\,Myr (top to bottom).}
\label{fig:valid_cmd} 
\end{figure}

Finally, Fig.~\ref{fig:valid_cmd} shows the (\ri)/\sloanr\ plane, which can be used to trace the ages and masses in a way similar to a Hertzsprung-Russell diagram. The objects shown in this plane have been dereddened according to their individual extinction estimates, such that we can compare their position against the theoretical isochrones (dashed lines) and evolutionary mass tracks (dotted lines) from the \citeauthor{siess} model. We find a good agreement between the position of objects in the diagram and their estimated masses. An equally good agreement is found for the age estimates (not shown).

\subsection{Comparison with existing spectroscopy}
\begin{figure}
\includegraphics[width=\linewidth]{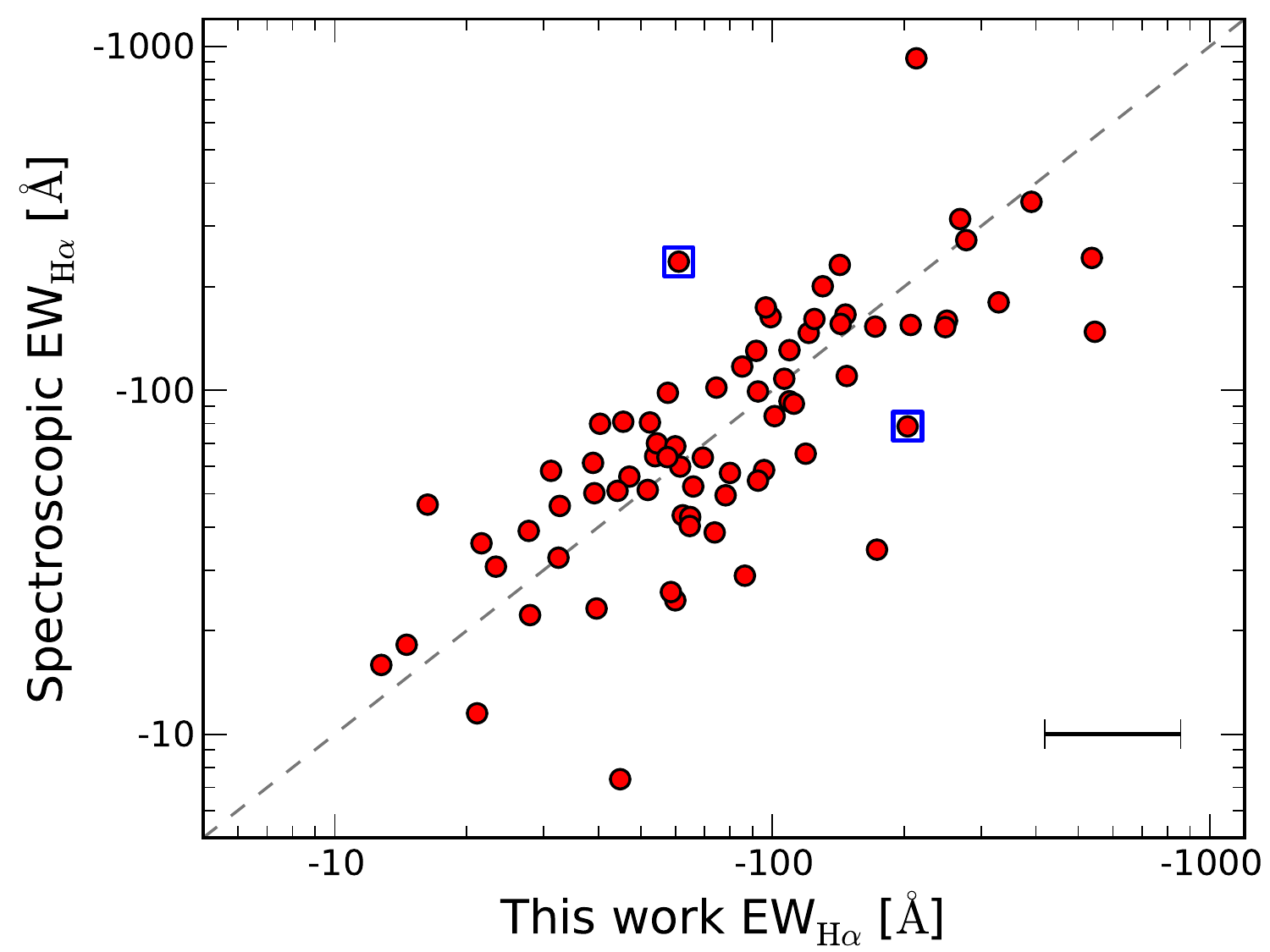}
\caption{Comparison of our inferred \halpha\ EWs with values obtained from grism spectroscopy by \citet{dahm2005}. 
The grey dashed line shows the unity relation. 
The median error bar is shown in the bottom right. 
The scatter is thought to be due to a combination of uncertainty and natural \halpha\ emission variability.}
\label{fig:ew_valid} 
\end{figure}
The most comprehensive set of spectroscopy towards NGC\,2264 that is currently available in the literature is the survey presented by \citet{dahm2005}, which is based on the Wide-Field Grism Spectrograph (WFGS) on the University of Hawaii 2.2\,m telescope on Mauna Kea, augmented with spectra from the Gemini North 8.2\,m telescope.

Their dataset provides spectroscopic \halpha\ EWs for 74 out of the 115 accreting objects in our sample. We find a good correlation between their values and our estimates (r=0.8) which is shown in Fig.~\ref{fig:ew_valid}, albeit with a significant scatter. The spread in the relationship is very similar to the one we previously found in a different cluster (Paper~I), and is thought to be due to a combination of natural variability and uncertainty (the median 1-sigma uncertainty for our estimates is shown in the bottom right of the plot).

We note that the work by \citeauthor{dahm2005} highlighted the strong natural variability of the \halpha\ line. Using spectra from multiple epochs between 1990 and 2003, the authors reported that nearly all of the CTTS (90\%) exhibited changes in the EW at or above the 10\% level, while 57\% varied at 50\% or greater. This confirms that the scatter is at least in part due to natural variability.

\subsection{Objects with low likelihoods}
\label{sec:lowlikelihood}
An advantage of the Bayesian method is that we may evaluate how well the model matches different objects by comparing their mean likelihood values (obtained from Eqn.~\ref{d2}). Using this information, we found a small number of $\sim$10 objects with significantly lower likelihoods than the main locus of stars. These 10 ``worst-fit'' objects have been marked by blue rectangles in Figs.~\ref{fig:valid_ccd}-\ref{fig:ew_valid} (object ID's: C13507, C22501, C27963, C28541, C31190, C31352, C33877, C36198, C36493, C37366).  

The markers reveal that several of these objects show strong \halpha\ emission in Fig.~\ref{fig:valid_ccd}, while the colours appear anomalous in  Fig.~\ref{fig:valid_ccd_reddening}, where they fall beyond the extreme blue edge of the locus. Likewise, a few fall below the main locus in Fig.~\ref{fig:valid_cmd}. 

We can think of four reasons to explain their position:
\begin{enumerate}
\item The objects might be blue \halpha-emitters in the background, e.g. interacting binaries, Be stars or unresolved planetary nebulae \citep{corradi2008} not related to NGC\,2264.
\item If the objects are genuine members, the presence of strong \halpha\ emission suggests that they are undergoing high levels of mass accretion, which is known to be a cause for continuum veiling in the red part of the spectrum. The origin of such emission is unclear however \citep{fischer2011}.
\item The objects might be affected by anomalous extinction. Three of the stars have previously been discussed by \citet{sung2008,sung2009} who classified them as ``BMS'' (for Below the pre-Main Sequence) based on their outlier position in the colour-magnitude diagram. \citeauthor{sung2008} supported the assumption that these are bona-fide young stars with a nearly edge-on disk. The dust grains in a disk tend to be larger than those in the interstellar medium, and hence the extinction law may differ significantly.
\item It is possible that the edge of the disk contaminates the colours of the star, depending on the inner radius, inclination and shape of the disk.
\end{enumerate}

If these outlier objects are genuine members, they provide evidence that our results would profit from a more advanced pre-main sequence model which includes the effects of continuum veiling due to accretion and dust in the circumstellar environment. 
We will discuss this prospect in \S6.

\subsection{The extinction prior}
\label{sec:extinctionprior}
We noted previously that the extinction distribution (Fig.~\ref{fig:hist}, panel c) shows a deficit of highly reddened objects when compared to the prior (blue solid line). The prior was chosen based on the distribution of spectroscopy-based extinctions determined by \citet{rebull2002} for 202 candidate members (\S2). 

The mismatch between our results and the prior is explained by the use of different selection criteria in both studies. On one hand, the membership selection by \citeauthor{rebull2002} made use of a combination of colour-magnitude diagrams and photometric variability, which is prone to the inclusion of background objects (i.e. out of the 17 objects for which the authors reported extinctions larger than $\av > 3$, only 2 passed our membership criteria). On the other hand, the use of optical photometry and X-ray observations in our criteria is likely to introduce a bias against highly reddened objects.

The mismatch illustrates that the extinction prior does not determine the results alone, but merely helps the data to constrain the parameters using the additional knowledge which we chose to include. To understand the influence of the prior, we repeated the inference procedure using a log-uniform prior P($\log$\a0) $\sim \mathcal{U}(-1, 1)$, which is less informed. We found this prior to produce near-identical results with mean $\log A_0=-0.32$ ($A_0=0.48$), which differs from the original mean extinction by only +0.05\,mag. The influence of this prior on the other parameters was found to be negligible, with mass estimates showing a median shift of +0.01\,\msol\ and the median age remaining constant.

We also repeated the experiment using the absolute uniform prior P(\a0) $\sim \mathcal{U}(0, 10)$, which favours high extinctions far more strongly than the log-uniform prior. This was found to change the extinction of individual objects by a factor 2.1 on average, hence raising the mean extinction of the sample to $\log A_0=-0.03$ ($A_0=0.93$). As a result, masses experienced a median shift of +0.11\,\msol\ and the median age increased from 3.0\,Myr to 4.3\,Myr.

The factor $\sim$2 corresponds to the typical 1-$\sigma$ uncertainty in the extinction of highly reddened objects in our results (e.g. Fig.~\ref{example3}). We therefore estimate that this is the level at which our method is able to constrain the extinction in regions where no prior information is available. The ability to constrain the extinction within a factor $\sim$2 meets the level of accuracy we may reasonably expect from the combination of \sloanr/\sloani/J magnitudes, and represents a significant improvement over the widely used practice of assuming a fixed extinction value in the absence of a spectroscopic data. Moreover, we are confident that including additional photometry at longer wavelengths (e.g. 2MASS H/K) can further strengthen our handle on the extinction in future work.

\section{Future extensions}

We envisage extending the method presented in this work on two fronts: (i) more comprehensive modelling of the individual stars, and (ii) modelling the global properties of the cluster.

First, in the previous section we found indications that our results would profit from a more detailed model of T Tauri stars, which should include the effects of continuum veiling due to accretion shocks, as well as the presence of dust in the circumstellar environment. 
We envisage employing a grid of radiation transfer models for this purpose, such as the widely used models by \citet{robitaille2006,robitaille2007}. 
We note that a maximum-likelihood fitting tool already exists for these models\footnote{http://www.astro.wisc.edu/protostars}. At present the tool only links observations to `best-fit' parameters however, and it does not reveal the full posterior.
In turn, a more detailed model invites the inclusion of photometry across a wider wavelength range. 
We note the possibility to include U- and g-band magnitudes from the UVEX and VPHAS galactic plane surveys \citep{uvex}, deep JHK-magnitudes from the UKIDSS surveys \citep{lucas2008} and infrared photometry from space-based telescopes. 

Second, in \S4 we explained that our findings on the global properties of NGC\,2264 must be interpreted with caution, because we did not incorporate the cluster properties as part of our probabilistic model. This would be desirable, because there are pertinent open issues in the current literature which require a careful treatment of the parameter uncertainties \citep[these questions include the ages of clusters, their age spreads, and the dependency of accretion rates on stellar masses, e.g. see][]{clarke2006,jeffries2011}. These questions may be addressed by adding the relevant cluster parameters to the top of the hierarchical model in Fig.~\ref{dag}.

It is worth emphasising that the goal of understanding clusters can be regarded as the problem of finding a hierarchical model which best explains the observations.
For this reason, we encourage others to adopt graphical Bayesian models as a generic framework to link observations to theory.

\section{Conclusions}
We showed how the theory of graphical Bayesian networks can be used to define a probabilistic model which allows extinction, age, mass and accretion rate to be inferred from IPHAS \sloanr/\halpha/\sloani\ and 2MASS J-band photometry without the need for spectroscopy. The model combined the evolutionary model due to \citet{siess} and the simulated photometry for \halpha\ emission-line stars due to \citet{barentsen2011} to compute probabilistic posterior distributions.

Compared to more popular plane-fitting or maximum-likelihood techniques, the advantages of our approach are that (i) we dealt with the degeneracy between stellar mass and extinction by considering the full probability distribution of solutions and (ii) we obtained meaningful expectation values and uncertainties by marginalising over nuisance parameters such as the distance and disc truncation radius. 

We used the Python/PyMC library to compute the model using a Markov Chain Monte Carlo (MCMC) algorithm, which was found to take only a small amount of programming effort (Appendix~A). We then applied the method to 587 low-mass members of the NGC\,2264 star-forming region and found a good agreement between our results and the position of stars in colour/magnitude diagrams, as well as literature spectroscopy. We performed an initial inspection of the sample and found that:
\begin{enumerate}
\item NGC\,2264 shows a median age of 3.0\,Myr, albeit with a large apparent dispersion between 1.8 and 4.5 Myr (25 and 75\% quartiles);
\item 115 objects ($20\pm2\%$) show fractional \halpha\ luminosities above the chromospheric saturation limit \citep[$\log L_{H\alpha}/L_* > -3.3$;][]{barrado2003} and are therefore very likely to be CTTS objects which are accreting gas from a circumstellar disc;
\item for these CTTS objects, we estimated mass accretion rates on the basis of \halpha\ luminosities and found them to range between $10^{-10.7}$ and $10^{-6.4}$\,\msol/yr with a median of $10^{-8.4}$\,\msol/yr.
\end{enumerate}

The results were shown to be consistent with existing spectroscopic studies in the literature. Our method achieved these results with great efficiency by depending only on photometry, and provides a significant step forward from previous photometric methods because our probabilistic approach ensures that nuisance parameters, such as extinction and distance, are fully included in the analysis with a clear picture on any degeneracies.

In future work, we envisage extending the method to include more physics. We note the possibility to utilise a grid of radiation transfer models which include the effects of continuum veiling and material in the circumstellar environment. In turn, our method would benefit from the inclusion of additional photometric bands across a wider wavelength range.

Graphical Bayesian models provide a generic framework for estimating parameters from sparse data. We expect that the approach will become increasingly important as a tool for the effective utilisation of large surveys, in particular once distance estimates from Gaia can be included. We remind the reader that our source code is made available online\footnote{https://github.com/barentsen} and encourage others to reuse or improve the code.

\section*{Acknowledgements}
We thank Nick Wright, Eduardo Martin, Eli Bressert, Christian Knigge, Roberto Raddi, Renko and Thomas Barclay for constructive comments which helped to improve the paper. We thank David W. Hogg for getting us to adopt probabilistic inference through his blog and talks. We thank the referee, Fr\'ed\'eric J. Pont, for insightful corrections which strengthened the paper.

Armagh Observatory is funded by major grants from the Department of Culture, Arts and Leisure (DCAL) for Northern Ireland and the UK Science and Technology Facilities Council (STFC).
The IPHAS survey is carried out at the Isaac Newton Telescope (INT), operated by the Isaac Newton Group in the Spanish Observatorio del Roque de los Muchachos of the Instituto de Astrofisica de Canarias. IPHAS data were processed by the Cambridge Astronomical Survey Unit at the Institute of Astronomy in Cambridge.
2MASS is a joint project of the University of Massachusetts \& IPAC/Caltech, funded by NASA/NSF.

Our work made use of TopCat \citep{topcat}, Aladin, PostgreSQL, Python (PyMC, PyFITS, APLpy modules) and Amazon EC2.
Mosaics were generated using the Montage software maintained by NASA/IPAC.

\label{lastpage}

\bsp

\appendix

\onecolumn
\section{Python source code for the inference model}
\footnotesize

\makeatletter
\def\PY@reset{\let\PY@it=\relax \let\PY@bf=\relax%
    \let\PY@ul=\relax \let\PY@tc=\relax%
    \let\PY@bc=\relax \let\PY@ff=\relax}
\def\PY@tok#1{\csname PY@tok@#1\endcsname}
\def\PY@toks#1+{\ifx\relax#1\empty\else%
    \PY@tok{#1}\expandafter\PY@toks\fi}
\def\PY@do#1{\PY@bc{\PY@tc{\PY@ul{%
    \PY@it{\PY@bf{\PY@ff{#1}}}}}}}
\def\PY#1#2{\PY@reset\PY@toks#1+\relax+\PY@do{#2}}

\def\PY@tok@gd{\def\PY@tc##1{\textcolor[rgb]{0.63,0.00,0.00}{##1}}}
\def\PY@tok@gu{\let\PY@bf=\textbf\def\PY@tc##1{\textcolor[rgb]{0.50,0.00,0.50}{##1}}}
\def\PY@tok@gt{\def\PY@tc##1{\textcolor[rgb]{0.00,0.25,0.82}{##1}}}
\def\PY@tok@gs{\let\PY@bf=\textbf}
\def\PY@tok@gr{\def\PY@tc##1{\textcolor[rgb]{1.00,0.00,0.00}{##1}}}
\def\PY@tok@cm{\let\PY@it=\textit\def\PY@tc##1{\textcolor[rgb]{0.25,0.50,0.50}{##1}}}
\def\PY@tok@vg{\def\PY@tc##1{\textcolor[rgb]{0.10,0.09,0.49}{##1}}}
\def\PY@tok@m{\def\PY@tc##1{\textcolor[rgb]{0.40,0.40,0.40}{##1}}}
\def\PY@tok@mh{\def\PY@tc##1{\textcolor[rgb]{0.40,0.40,0.40}{##1}}}
\def\PY@tok@go{\def\PY@tc##1{\textcolor[rgb]{0.50,0.50,0.50}{##1}}}
\def\PY@tok@ge{\let\PY@it=\textit}
\def\PY@tok@vc{\def\PY@tc##1{\textcolor[rgb]{0.10,0.09,0.49}{##1}}}
\def\PY@tok@il{\def\PY@tc##1{\textcolor[rgb]{0.40,0.40,0.40}{##1}}}
\def\PY@tok@cs{\let\PY@it=\textit\def\PY@tc##1{\textcolor[rgb]{0.25,0.50,0.50}{##1}}}
\def\PY@tok@cp{\def\PY@tc##1{\textcolor[rgb]{0.74,0.48,0.00}{##1}}}
\def\PY@tok@gi{\def\PY@tc##1{\textcolor[rgb]{0.00,0.63,0.00}{##1}}}
\def\PY@tok@gh{\let\PY@bf=\textbf\def\PY@tc##1{\textcolor[rgb]{0.00,0.00,0.50}{##1}}}
\def\PY@tok@ni{\let\PY@bf=\textbf\def\PY@tc##1{\textcolor[rgb]{0.60,0.60,0.60}{##1}}}
\def\PY@tok@nl{\def\PY@tc##1{\textcolor[rgb]{0.63,0.63,0.00}{##1}}}
\def\PY@tok@nn{\let\PY@bf=\textbf\def\PY@tc##1{\textcolor[rgb]{0.00,0.00,1.00}{##1}}}
\def\PY@tok@no{\def\PY@tc##1{\textcolor[rgb]{0.53,0.00,0.00}{##1}}}
\def\PY@tok@na{\def\PY@tc##1{\textcolor[rgb]{0.49,0.56,0.16}{##1}}}
\def\PY@tok@nb{\def\PY@tc##1{\textcolor[rgb]{0.00,0.50,0.00}{##1}}}
\def\PY@tok@nc{\let\PY@bf=\textbf\def\PY@tc##1{\textcolor[rgb]{0.00,0.00,1.00}{##1}}}
\def\PY@tok@nd{\def\PY@tc##1{\textcolor[rgb]{0.67,0.13,1.00}{##1}}}
\def\PY@tok@ne{\let\PY@bf=\textbf\def\PY@tc##1{\textcolor[rgb]{0.82,0.25,0.23}{##1}}}
\def\PY@tok@nf{\def\PY@tc##1{\textcolor[rgb]{0.00,0.00,1.00}{##1}}}
\def\PY@tok@si{\let\PY@bf=\textbf\def\PY@tc##1{\textcolor[rgb]{0.73,0.40,0.53}{##1}}}
\def\PY@tok@s2{\def\PY@tc##1{\textcolor[rgb]{0.73,0.13,0.13}{##1}}}
\def\PY@tok@vi{\def\PY@tc##1{\textcolor[rgb]{0.10,0.09,0.49}{##1}}}
\def\PY@tok@nt{\let\PY@bf=\textbf\def\PY@tc##1{\textcolor[rgb]{0.00,0.50,0.00}{##1}}}
\def\PY@tok@nv{\def\PY@tc##1{\textcolor[rgb]{0.10,0.09,0.49}{##1}}}
\def\PY@tok@s1{\def\PY@tc##1{\textcolor[rgb]{0.73,0.13,0.13}{##1}}}
\def\PY@tok@sh{\def\PY@tc##1{\textcolor[rgb]{0.73,0.13,0.13}{##1}}}
\def\PY@tok@sc{\def\PY@tc##1{\textcolor[rgb]{0.73,0.13,0.13}{##1}}}
\def\PY@tok@sx{\def\PY@tc##1{\textcolor[rgb]{0.00,0.50,0.00}{##1}}}
\def\PY@tok@bp{\def\PY@tc##1{\textcolor[rgb]{0.00,0.50,0.00}{##1}}}
\def\PY@tok@c1{\let\PY@it=\textit\def\PY@tc##1{\textcolor[rgb]{0.25,0.50,0.50}{##1}}}
\def\PY@tok@kc{\let\PY@bf=\textbf\def\PY@tc##1{\textcolor[rgb]{0.00,0.50,0.00}{##1}}}
\def\PY@tok@c{\let\PY@it=\textit\def\PY@tc##1{\textcolor[rgb]{0.25,0.50,0.50}{##1}}}
\def\PY@tok@mf{\def\PY@tc##1{\textcolor[rgb]{0.40,0.40,0.40}{##1}}}
\def\PY@tok@err{\def\PY@bc##1{\fcolorbox[rgb]{1.00,0.00,0.00}{1,1,1}{##1}}}
\def\PY@tok@kd{\let\PY@bf=\textbf\def\PY@tc##1{\textcolor[rgb]{0.00,0.50,0.00}{##1}}}
\def\PY@tok@ss{\def\PY@tc##1{\textcolor[rgb]{0.10,0.09,0.49}{##1}}}
\def\PY@tok@sr{\def\PY@tc##1{\textcolor[rgb]{0.73,0.40,0.53}{##1}}}
\def\PY@tok@mo{\def\PY@tc##1{\textcolor[rgb]{0.40,0.40,0.40}{##1}}}
\def\PY@tok@kn{\let\PY@bf=\textbf\def\PY@tc##1{\textcolor[rgb]{0.00,0.50,0.00}{##1}}}
\def\PY@tok@mi{\def\PY@tc##1{\textcolor[rgb]{0.40,0.40,0.40}{##1}}}
\def\PY@tok@gp{\let\PY@bf=\textbf\def\PY@tc##1{\textcolor[rgb]{0.00,0.00,0.50}{##1}}}
\def\PY@tok@o{\def\PY@tc##1{\textcolor[rgb]{0.40,0.40,0.40}{##1}}}
\def\PY@tok@kr{\let\PY@bf=\textbf\def\PY@tc##1{\textcolor[rgb]{0.00,0.50,0.00}{##1}}}
\def\PY@tok@s{\def\PY@tc##1{\textcolor[rgb]{0.73,0.13,0.13}{##1}}}
\def\PY@tok@kp{\def\PY@tc##1{\textcolor[rgb]{0.00,0.50,0.00}{##1}}}
\def\PY@tok@w{\def\PY@tc##1{\textcolor[rgb]{0.73,0.73,0.73}{##1}}}
\def\PY@tok@kt{\def\PY@tc##1{\textcolor[rgb]{0.69,0.00,0.25}{##1}}}
\def\PY@tok@ow{\let\PY@bf=\textbf\def\PY@tc##1{\textcolor[rgb]{0.67,0.13,1.00}{##1}}}
\def\PY@tok@sb{\def\PY@tc##1{\textcolor[rgb]{0.73,0.13,0.13}{##1}}}
\def\PY@tok@k{\let\PY@bf=\textbf\def\PY@tc##1{\textcolor[rgb]{0.00,0.50,0.00}{##1}}}
\def\PY@tok@se{\let\PY@bf=\textbf\def\PY@tc##1{\textcolor[rgb]{0.73,0.40,0.13}{##1}}}
\def\PY@tok@sd{\let\PY@it=\textit\def\PY@tc##1{\textcolor[rgb]{0.73,0.13,0.13}{##1}}}

\def\PYZbs{\char`\\}
\def\PYZus{\char`\_}
\def\PYZob{\char`\{}
\def\PYZcb{\char`\}}
\def\PYZca{\char`\^}
\def\PYZsh{\char`\#}
\def\PYZpc{\char`\%}
\def\PYZdl{\char`\$}
\def\PYZti{\char`\~}
\def\PYZat{@}
\def\PYZlb{[}
\def\PYZrb{]}
\makeatother

\begin{Verbatim}[commandchars=\\\{\}]
\PY{k+kn}{import} \PY{n+nn}{numpy} \PY{k+kn}{as} \PY{n+nn}{np}
\PY{k+kn}{from} \PY{n+nn}{scipy.interpolate.rbf} \PY{k+kn}{import} \PY{n}{Rbf}
\PY{k+kn}{import} \PY{n+nn}{pyfits}
\PY{k+kn}{import} \PY{n+nn}{pymc}

\PY{l+s+sd}{""" Interpolation functions for intrinsic magnitudes """}
\PY{n}{siess} \PY{o}{=} \PY{n}{pyfits}\PY{o}{.}\PY{n}{getdata}\PY{p}{(}\PY{l+s}{"}\PY{l+s}{siess\PYZus{}isochrones.fits}\PY{l+s}{"}\PY{p}{,} \PY{l+m+mi}{1}\PY{p}{)}  \PY{c}{\PYZsh{} Siess et al. (2000)}
\PY{c}{\PYZsh{} Interpolation is performed using linear Radial Basis Functions}
\PY{n}{siess\PYZus{}Mr} \PY{o}{=} \PY{n}{Rbf}\PY{p}{(}\PY{n}{siess}\PY{o}{.}\PY{n}{field}\PY{p}{(}\PY{l+s}{"}\PY{l+s}{logMass}\PY{l+s}{"}\PY{p}{)}\PY{p}{,} \PY{n}{siess}\PY{o}{.}\PY{n}{field}\PY{p}{(}\PY{l+s}{"}\PY{l+s}{logAge}\PY{l+s}{"}\PY{p}{)}\PY{p}{,}
               \PY{n}{siess}\PY{o}{.}\PY{n}{field}\PY{p}{(}\PY{l+s}{"}\PY{l+s}{Mr\PYZus{}iphas}\PY{l+s}{"}\PY{p}{)}\PY{p}{,} \PY{n}{function}\PY{o}{=}\PY{l+s}{"}\PY{l+s}{linear}\PY{l+s}{"}\PY{p}{)}
\PY{n}{siess\PYZus{}Mi} \PY{o}{=} \PY{n}{Rbf}\PY{p}{(}\PY{n}{siess}\PY{o}{.}\PY{n}{field}\PY{p}{(}\PY{l+s}{"}\PY{l+s}{logMass}\PY{l+s}{"}\PY{p}{)}\PY{p}{,} \PY{n}{siess}\PY{o}{.}\PY{n}{field}\PY{p}{(}\PY{l+s}{"}\PY{l+s}{logAge}\PY{l+s}{"}\PY{p}{)}\PY{p}{,}
               \PY{n}{siess}\PY{o}{.}\PY{n}{field}\PY{p}{(}\PY{l+s}{"}\PY{l+s}{Mi\PYZus{}iphas}\PY{l+s}{"}\PY{p}{)}\PY{p}{,} \PY{n}{function}\PY{o}{=}\PY{l+s}{"}\PY{l+s}{linear}\PY{l+s}{"}\PY{p}{)}
\PY{n}{siess\PYZus{}Mj} \PY{o}{=} \PY{n}{Rbf}\PY{p}{(}\PY{n}{siess}\PY{o}{.}\PY{n}{field}\PY{p}{(}\PY{l+s}{"}\PY{l+s}{logMass}\PY{l+s}{"}\PY{p}{)}\PY{p}{,} \PY{n}{siess}\PY{o}{.}\PY{n}{field}\PY{p}{(}\PY{l+s}{"}\PY{l+s}{logAge}\PY{l+s}{"}\PY{p}{)}\PY{p}{,}
               \PY{n}{siess}\PY{o}{.}\PY{n}{field}\PY{p}{(}\PY{l+s}{"}\PY{l+s}{Mj}\PY{l+s}{"}\PY{p}{)}\PY{p}{,} \PY{n}{function}\PY{o}{=}\PY{l+s}{"}\PY{l+s}{linear}\PY{l+s}{"}\PY{p}{)}
\PY{n}{siess\PYZus{}logR} \PY{o}{=} \PY{n}{Rbf}\PY{p}{(}\PY{n}{siess}\PY{o}{.}\PY{n}{field}\PY{p}{(}\PY{l+s}{"}\PY{l+s}{logMass}\PY{l+s}{"}\PY{p}{)}\PY{p}{,} \PY{n}{siess}\PY{o}{.}\PY{n}{field}\PY{p}{(}\PY{l+s}{"}\PY{l+s}{logAge}\PY{l+s}{"}\PY{p}{)}\PY{p}{,}
                 \PY{n}{siess}\PY{o}{.}\PY{n}{field}\PY{p}{(}\PY{l+s}{"}\PY{l+s}{logRadius}\PY{l+s}{"}\PY{p}{)}\PY{p}{,} \PY{n}{function}\PY{o}{=}\PY{l+s}{"}\PY{l+s}{linear}\PY{l+s}{"}\PY{p}{)}

\PY{l+s+sd}{""" Functions for magnitude offsets due to emission & exctinction """}
\PY{n}{sim} \PY{o}{=} \PY{n}{pyfits}\PY{o}{.}\PY{n}{getdata}\PY{p}{(}\PY{l+s}{"}\PY{l+s}{simulated\PYZus{}iphas\PYZus{}colours\PYZus{}barentsen2011.fits}\PY{l+s}{"}\PY{p}{,} \PY{l+m+mi}{1}\PY{p}{)}  \PY{c}{\PYZsh{} PaperI}
\PY{c}{\PYZsh{} Functions for r'/Ha/i' offsets as a function of colour, extinction and EW}
\PY{n}{r\PYZus{}offset} \PY{o}{=} \PY{n}{Rbf}\PY{p}{(}\PY{n}{sim}\PY{o}{.}\PY{n}{field}\PY{p}{(}\PY{l+s}{"}\PY{l+s}{ri\PYZus{}unred}\PY{l+s}{"}\PY{p}{)}\PY{p}{,} \PY{n}{sim}\PY{o}{.}\PY{n}{field}\PY{p}{(}\PY{l+s}{"}\PY{l+s}{av}\PY{l+s}{"}\PY{p}{)}\PY{p}{,} \PY{n}{sim}\PY{o}{.}\PY{n}{field}\PY{p}{(}\PY{l+s}{"}\PY{l+s}{logew}\PY{l+s}{"}\PY{p}{)}\PY{p}{,}
               \PY{n}{sim}\PY{o}{.}\PY{n}{field}\PY{p}{(}\PY{l+s}{"}\PY{l+s}{d\PYZus{}r}\PY{l+s}{"}\PY{p}{)}\PY{p}{,} \PY{n}{function}\PY{o}{=}\PY{l+s}{"}\PY{l+s}{linear}\PY{l+s}{"}\PY{p}{)}
\PY{n}{ha\PYZus{}offset} \PY{o}{=} \PY{n}{Rbf}\PY{p}{(}\PY{n}{sim}\PY{o}{.}\PY{n}{field}\PY{p}{(}\PY{l+s}{"}\PY{l+s}{ri\PYZus{}unred}\PY{l+s}{"}\PY{p}{)}\PY{p}{,} \PY{n}{sim}\PY{o}{.}\PY{n}{field}\PY{p}{(}\PY{l+s}{"}\PY{l+s}{av}\PY{l+s}{"}\PY{p}{)}\PY{p}{,} \PY{n}{sim}\PY{o}{.}\PY{n}{field}\PY{p}{(}\PY{l+s}{"}\PY{l+s}{logew}\PY{l+s}{"}\PY{p}{)}\PY{p}{,}
                \PY{n}{sim}\PY{o}{.}\PY{n}{field}\PY{p}{(}\PY{l+s}{"}\PY{l+s}{d\PYZus{}ha}\PY{l+s}{"}\PY{p}{)}\PY{p}{,} \PY{n}{function}\PY{o}{=}\PY{l+s}{"}\PY{l+s}{linear}\PY{l+s}{"}\PY{p}{)}
\PY{n}{i\PYZus{}offset} \PY{o}{=} \PY{n}{Rbf}\PY{p}{(}\PY{n}{sim}\PY{o}{.}\PY{n}{field}\PY{p}{(}\PY{l+s}{"}\PY{l+s}{ri\PYZus{}unred}\PY{l+s}{"}\PY{p}{)}\PY{p}{,} \PY{n}{sim}\PY{o}{.}\PY{n}{field}\PY{p}{(}\PY{l+s}{"}\PY{l+s}{av}\PY{l+s}{"}\PY{p}{)}\PY{p}{,} \PY{n}{sim}\PY{o}{.}\PY{n}{field}\PY{p}{(}\PY{l+s}{"}\PY{l+s}{logew}\PY{l+s}{"}\PY{p}{)}\PY{p}{,}
               \PY{n}{sim}\PY{o}{.}\PY{n}{field}\PY{p}{(}\PY{l+s}{"}\PY{l+s}{d\PYZus{}i}\PY{l+s}{"}\PY{p}{)}\PY{p}{,} \PY{n}{function}\PY{o}{=}\PY{l+s}{"}\PY{l+s}{linear}\PY{l+s}{"}\PY{p}{)}
\PY{c}{\PYZsh{} Intrinsic (r'-Ha) colour as a function of intrinsic (r'-i')}
\PY{n}{intrinsic} \PY{o}{=} \PY{p}{(}\PY{n}{sim}\PY{o}{.}\PY{n}{field}\PY{p}{(}\PY{l+s}{"}\PY{l+s}{av}\PY{l+s}{"}\PY{p}{)} \PY{o}{==} \PY{l+m+mi}{0}\PY{p}{)} \PY{o}{&} \PY{p}{(}\PY{n}{sim}\PY{o}{.}\PY{n}{field}\PY{p}{(}\PY{l+s}{"}\PY{l+s}{logew}\PY{l+s}{"}\PY{p}{)} \PY{o}{==} \PY{o}{-}\PY{l+m+mi}{1}\PY{p}{)}
\PY{n}{rminHa\PYZus{}intrinsic} \PY{o}{=} \PY{n}{Rbf}\PY{p}{(}\PY{n}{sim}\PY{o}{.}\PY{n}{field}\PY{p}{(}\PY{l+s}{"}\PY{l+s}{ri\PYZus{}unred}\PY{l+s}{"}\PY{p}{)}\PY{p}{[}\PY{n}{intrinsic}\PY{p}{]}\PY{p}{,}
                       \PY{n}{sim}\PY{o}{.}\PY{n}{field}\PY{p}{(}\PY{l+s}{"}\PY{l+s}{rha}\PY{l+s}{"}\PY{p}{)}\PY{p}{[}\PY{n}{intrinsic}\PY{p}{]}\PY{p}{,} \PY{n}{function}\PY{o}{=}\PY{l+s}{"}\PY{l+s}{linear}\PY{l+s}{"}\PY{p}{)}


\PY{k}{def} \PY{n+nf}{make\PYZus{}model}\PY{p}{(}\PY{n}{observed\PYZus{}sed}\PY{p}{,} \PY{n}{e\PYZus{}observed\PYZus{}sed}\PY{p}{)}\PY{p}{:}
    \PY{l+s+sd}{""" This function returns all prior and likelihood objects """}

    \PY{c}{\PYZsh{} Prior: mass (Kroupa 2001)}
    \PY{n+nd}{@pymc.stochastic}\PY{p}{(}\PY{p}{)}
    \PY{k}{def} \PY{n+nf}{logM}\PY{p}{(}\PY{n}{value}\PY{o}{=}\PY{n}{np}\PY{o}{.}\PY{n}{array}\PY{p}{(}\PY{p}{[}\PY{n}{np}\PY{o}{.}\PY{n}{log10}\PY{p}{(}\PY{l+m+mf}{0.5}\PY{p}{)}\PY{p}{]}\PY{p}{)}\PY{p}{,} \PY{n}{a}\PY{o}{=}\PY{n}{np}\PY{o}{.}\PY{n}{log10}\PY{p}{(}\PY{l+m+mf}{0.1}\PY{p}{)}\PY{p}{,} \PY{n}{b}\PY{o}{=}\PY{n}{np}\PY{o}{.}\PY{n}{log10}\PY{p}{(}\PY{l+m+mi}{7}\PY{p}{)}\PY{p}{)}\PY{p}{:}

        \PY{k}{def} \PY{n+nf}{logp}\PY{p}{(}\PY{n}{value}\PY{p}{,} \PY{n}{a}\PY{p}{,} \PY{n}{b}\PY{p}{)}\PY{p}{:}
            \PY{k}{if} \PY{n}{value} \PY{o}{>} \PY{n}{b} \PY{o+ow}{or} \PY{n}{value} \PY{o}{<} \PY{n}{a}\PY{p}{:}
                \PY{k}{return} \PY{o}{-}\PY{n}{np}\PY{o}{.}\PY{n}{Inf}  \PY{c}{\PYZsh{} Stay within the model limits (a,b).}
            \PY{k}{else}\PY{p}{:}
                \PY{n}{mass} \PY{o}{=} \PY{l+m+mi}{10} \PY{o}{*}\PY{o}{*} \PY{n}{value}
                \PY{k}{if} \PY{n}{mass} \PY{o}{<} \PY{l+m+mf}{0.5}\PY{p}{:}
                    \PY{k}{return} \PY{n}{np}\PY{o}{.}\PY{n}{log}\PY{p}{(}\PY{n}{mass} \PY{o}{*}\PY{o}{*} \PY{o}{-}\PY{l+m+mf}{1.3}\PY{p}{)}  \PY{c}{\PYZsh{} Kroupa (2001)}
                \PY{k}{else}\PY{p}{:}
                    \PY{k}{return} \PY{n}{np}\PY{o}{.}\PY{n}{log}\PY{p}{(}\PY{l+m+mf}{0.5} \PY{o}{*} \PY{n}{mass} \PY{o}{*}\PY{o}{*} \PY{o}{-}\PY{l+m+mf}{2.3}\PY{p}{)}  \PY{c}{\PYZsh{} Kroupa (2001)}

        \PY{k}{def} \PY{n+nf}{random}\PY{p}{(}\PY{n}{a}\PY{p}{,} \PY{n}{b}\PY{p}{)}\PY{p}{:}
            \PY{n}{val} \PY{o}{=} \PY{p}{(}\PY{n}{b} \PY{o}{-} \PY{n}{a}\PY{p}{)} \PY{o}{*} \PY{n}{np}\PY{o}{.}\PY{n}{random}\PY{o}{.}\PY{n}{rand}\PY{p}{(}\PY{p}{)} \PY{o}{+} \PY{n}{a}
            \PY{k}{return} \PY{n}{np}\PY{o}{.}\PY{n}{array}\PY{p}{(}\PY{p}{[}\PY{n}{val}\PY{p}{]}\PY{p}{)}

    \PY{c}{\PYZsh{} Prior: age (uniform in the logarithm)}
    \PY{n}{logT} \PY{o}{=} \PY{n}{pymc}\PY{o}{.}\PY{n}{Uniform}\PY{p}{(}\PY{l+s}{"}\PY{l+s}{logT}\PY{l+s}{"}\PY{p}{,} \PY{n}{np}\PY{o}{.}\PY{n}{array}\PY{p}{(}\PY{p}{[}\PY{l+m+mi}{5}\PY{p}{]}\PY{p}{)}\PY{p}{,} \PY{n}{np}\PY{o}{.}\PY{n}{array}\PY{p}{(}\PY{p}{[}\PY{l+m+mi}{8}\PY{p}{]}\PY{p}{)}\PY{p}{)}

    \PY{c}{\PYZsh{} Prior: accretion rate (uniform in the logarithm)}
    \PY{n}{logMacc} \PY{o}{=} \PY{n}{pymc}\PY{o}{.}\PY{n}{Uniform}\PY{p}{(}\PY{l+s}{"}\PY{l+s}{logMacc}\PY{l+s}{"}\PY{p}{,} \PY{n}{np}\PY{o}{.}\PY{n}{array}\PY{p}{(}\PY{p}{[}\PY{o}{-}\PY{l+m+mi}{15}\PY{p}{]}\PY{p}{)}\PY{p}{,} \PY{n}{np}\PY{o}{.}\PY{n}{array}\PY{p}{(}\PY{p}{[}\PY{o}{-}\PY{l+m+mi}{2}\PY{p}{]}\PY{p}{)}\PY{p}{)}

    \PY{c}{\PYZsh{} Prior: disc truncation radius (Rin = 5 +\PYZbs{}- 2 R, Gullbring et al. 1998)}
    \PY{n}{Rin} \PY{o}{=} \PY{n}{pymc}\PY{o}{.}\PY{n}{TruncatedNormal}\PY{p}{(}\PY{l+s}{"}\PY{l+s}{Rin}\PY{l+s}{"}\PY{p}{,} \PY{n}{mu}\PY{o}{=}\PY{n}{np}\PY{o}{.}\PY{n}{array}\PY{p}{(}\PY{p}{[}\PY{l+m+mf}{5.0}\PY{p}{]}\PY{p}{)}\PY{p}{,} \PY{n}{tau}\PY{o}{=}\PY{l+m+mf}{2.0} \PY{o}{*}\PY{o}{*} \PY{o}{-}\PY{l+m+mi}{2}\PY{p}{,}
                               \PY{n}{a}\PY{o}{=}\PY{l+m+mf}{1.01}\PY{p}{,} \PY{n}{b}\PY{o}{=}\PY{l+m+mf}{9e99}\PY{p}{)}

    \PY{c}{\PYZsh{} Prior: distance (d = 760 +\PYZbs{}- 5 pc, Sung 1997)}
    \PY{n}{d} \PY{o}{=} \PY{n}{pymc}\PY{o}{.}\PY{n}{TruncatedNormal}\PY{p}{(}\PY{l+s}{"}\PY{l+s}{d}\PY{l+s}{"}\PY{p}{,} \PY{n}{mu}\PY{o}{=}\PY{n}{np}\PY{o}{.}\PY{n}{array}\PY{p}{(}\PY{p}{[}\PY{l+m+mf}{760.0}\PY{p}{]}\PY{p}{)}\PY{p}{,} \PY{n}{tau}\PY{o}{=}\PY{l+m+mf}{5.0} \PY{o}{*}\PY{o}{*} \PY{o}{-}\PY{l+m+mi}{2}\PY{p}{,}
                             \PY{n}{a}\PY{o}{=}\PY{l+m+mi}{700}\PY{p}{,} \PY{n}{b}\PY{o}{=}\PY{l+m+mf}{9e99}\PY{p}{)}

    \PY{c}{\PYZsh{} Prior: extinction (logA0 = -0.27 +/- 0.46, Rebull et al. 2002)}
    \PY{n}{logA0} \PY{o}{=} \PY{n}{pymc}\PY{o}{.}\PY{n}{Normal}\PY{p}{(}\PY{l+s}{"}\PY{l+s}{logA0}\PY{l+s}{"}\PY{p}{,} \PY{n}{mu}\PY{o}{=}\PY{n}{np}\PY{o}{.}\PY{n}{array}\PY{p}{(}\PY{p}{[}\PY{o}{-}\PY{l+m+mf}{0.27}\PY{p}{]}\PY{p}{)}\PY{p}{,} \PY{n}{tau}\PY{o}{=}\PY{l+m+mf}{0.46} \PY{o}{*}\PY{o}{*} \PY{o}{-}\PY{l+m+mi}{2}\PY{p}{)}

    \PY{c}{\PYZsh{} Likelihood: intrinsic SED}
    \PY{n+nd}{@pymc.deterministic}\PY{p}{(}\PY{p}{)}
    \PY{k}{def} \PY{n+nf}{SED\PYZus{}intrinsic}\PY{p}{(}\PY{n}{logM}\PY{o}{=}\PY{n}{logM}\PY{p}{,} \PY{n}{logT}\PY{o}{=}\PY{n}{logT}\PY{p}{)}\PY{p}{:}
        \PY{n}{r} \PY{o}{=} \PY{n}{siess\PYZus{}Mr}\PY{p}{(}\PY{n}{logM}\PY{p}{,} \PY{n}{logT}\PY{p}{)}  \PY{c}{\PYZsh{} IPHAS r' as a function of (mass, age)}
        \PY{n}{i} \PY{o}{=} \PY{n}{siess\PYZus{}Mi}\PY{p}{(}\PY{n}{logM}\PY{p}{,} \PY{n}{logT}\PY{p}{)}  \PY{c}{\PYZsh{} IPHAS i}
        \PY{n}{j} \PY{o}{=} \PY{n}{siess\PYZus{}Mj}\PY{p}{(}\PY{n}{logM}\PY{p}{,} \PY{n}{logT}\PY{p}{)}  \PY{c}{\PYZsh{} 2MASS J}
        \PY{n}{ha} \PY{o}{=} \PY{n}{r} \PY{o}{-} \PY{n}{rminHa\PYZus{}intrinsic}\PY{p}{(}\PY{n}{r} \PY{o}{-} \PY{n}{i}\PY{p}{)}  \PY{c}{\PYZsh{} IPHAS H-alpha}
        \PY{k}{return} \PY{n}{np}\PY{o}{.}\PY{n}{array}\PY{p}{(}\PY{p}{[}\PY{n}{r}\PY{p}{[}\PY{l+m+mi}{0}\PY{p}{]}\PY{p}{,} \PY{n}{ha}\PY{p}{[}\PY{l+m+mi}{0}\PY{p}{]}\PY{p}{,} \PY{n}{i}\PY{p}{[}\PY{l+m+mi}{0}\PY{p}{]}\PY{p}{,} \PY{n}{j}\PY{p}{[}\PY{l+m+mi}{0}\PY{p}{]}\PY{p}{]}\PY{p}{)}

    \PY{c}{\PYZsh{} Likelihood: H-alpha excess luminosity}
    \PY{n+nd}{@pymc.deterministic}\PY{p}{(}\PY{p}{)}
    \PY{k}{def} \PY{n+nf}{logLacc}\PY{p}{(}\PY{n}{logM}\PY{o}{=}\PY{n}{logM}\PY{p}{,} \PY{n}{logT}\PY{o}{=}\PY{n}{logT}\PY{p}{,} \PY{n}{logMacc}\PY{o}{=}\PY{n}{logMacc}\PY{p}{,} \PY{n}{Rin}\PY{o}{=}\PY{n}{Rin}\PY{p}{)}\PY{p}{:}
        \PY{n}{logR} \PY{o}{=} \PY{n}{siess\PYZus{}logR}\PY{p}{(}\PY{n}{logM}\PY{p}{,} \PY{n}{logT}\PY{p}{)}  \PY{c}{\PYZsh{} Radius as a function of (mass, age)}
        \PY{k}{return} \PY{l+m+mf}{7.496} \PY{o}{+} \PY{n}{logM} \PY{o}{+} \PY{n}{logMacc} \PY{o}{-} \PY{n}{logR} \PY{o}{+} \PY{n}{np}\PY{o}{.}\PY{n}{log10}\PY{p}{(}\PY{l+m+mi}{1} \PY{o}{-} \PY{p}{(}\PY{l+m+mi}{1} \PY{o}{/} \PY{n}{Rin}\PY{p}{)}\PY{p}{)}
    \PY{n}{logLha} \PY{o}{=} \PY{n}{pymc}\PY{o}{.}\PY{n}{Normal}\PY{p}{(}\PY{l+s}{"}\PY{l+s}{logLha}\PY{l+s}{"}\PY{p}{,} \PY{n}{mu}\PY{o}{=}\PY{p}{(}\PY{l+m+mf}{0.64} \PY{o}{*} \PY{n}{logLacc} \PY{o}{-} \PY{l+m+mf}{2.12}\PY{p}{)}\PY{p}{,} \PY{n}{tau}\PY{o}{=}\PY{l+m+mf}{0.43} \PY{o}{*}\PY{o}{*} \PY{o}{-}\PY{l+m+mi}{2}\PY{p}{)}

    \PY{c}{\PYZsh{} Likelihood: H-alpha equivalent width (EW).}
    \PY{n+nd}{@pymc.deterministic}\PY{p}{(}\PY{p}{)}
    \PY{k}{def} \PY{n+nf}{logEW}\PY{p}{(}\PY{n}{logLha}\PY{o}{=}\PY{n}{logLha}\PY{p}{,} \PY{n}{SED\PYZus{}intrinsic}\PY{o}{=}\PY{n}{SED\PYZus{}intrinsic}\PY{p}{)}\PY{p}{:}
        \PY{n}{Lha} \PY{o}{=} \PY{l+m+mi}{10} \PY{o}{*}\PY{o}{*} \PY{n}{logLha}  \PY{c}{\PYZsh{} Excess luminosity}
        \PY{n}{Lha\PYZus{}con} \PY{o}{=} \PY{l+m+mf}{0.316} \PY{o}{*} \PY{l+m+mi}{10} \PY{o}{*}\PY{o}{*} \PY{p}{(}\PY{o}{-}\PY{l+m+mf}{0.4} \PY{o}{*} \PY{p}{(}\PY{n}{SED\PYZus{}intrinsic}\PY{p}{[}\PY{l+m+mi}{1}\PY{p}{]} \PY{o}{+} \PY{l+m+mf}{0.03}\PY{p}{)}\PY{p}{)}  \PY{c}{\PYZsh{} Continuum}
        \PY{n}{ew} \PY{o}{=} \PY{o}{-}\PY{l+m+mf}{95.0} \PY{o}{*} \PY{n}{Lha} \PY{o}{/} \PY{n}{Lha\PYZus{}con}  \PY{c}{\PYZsh{} Equivalent width.}
        \PY{k}{return} \PY{n}{np}\PY{o}{.}\PY{n}{log10}\PY{p}{(}\PY{o}{-}\PY{n}{ew}\PY{p}{)}

    \PY{c}{\PYZsh{} Likelihood: apparent SED}
    \PY{n+nd}{@pymc.deterministic}\PY{p}{(}\PY{p}{)}
    \PY{k}{def} \PY{n+nf}{SED\PYZus{}apparent}\PY{p}{(}\PY{n}{d}\PY{o}{=}\PY{n}{d}\PY{p}{,} \PY{n}{logA0}\PY{o}{=}\PY{n}{logA0}\PY{p}{,} \PY{n}{SED\PYZus{}intr}\PY{o}{=}\PY{n}{SED\PYZus{}intrinsic}\PY{p}{,} \PY{n}{logEW}\PY{o}{=}\PY{n}{logEW}\PY{p}{)}\PY{p}{:}
        \PY{n}{dismod} \PY{o}{=} \PY{l+m+mf}{5.0} \PY{o}{*} \PY{n}{np}\PY{o}{.}\PY{n}{log10}\PY{p}{(}\PY{n}{d}\PY{p}{)} \PY{o}{-} \PY{l+m+mf}{5.0}  \PY{c}{\PYZsh{} Distance modulus.}
        \PY{n}{A0} \PY{o}{=} \PY{l+m+mf}{10.0} \PY{o}{*}\PY{o}{*} \PY{n}{logA0}  \PY{c}{\PYZsh{} Extinction parameter}
        \PY{n}{ri\PYZus{}intr} \PY{o}{=} \PY{n}{np}\PY{o}{.}\PY{n}{array}\PY{p}{(}\PY{p}{[}\PY{n}{SED\PYZus{}intr}\PY{p}{[}\PY{l+m+mi}{0}\PY{p}{]} \PY{o}{-} \PY{n}{SED\PYZus{}intr}\PY{p}{[}\PY{l+m+mi}{2}\PY{p}{]}\PY{p}{]}\PY{p}{)}  \PY{c}{\PYZsh{} Intrinsic (r'-i')}
        \PY{c}{\PYZsh{} Correct the intrinsic magnitudes for extinction and H-alpha emission:}
        \PY{n}{r} \PY{o}{=} \PY{n}{SED\PYZus{}intr}\PY{p}{[}\PY{l+m+mi}{0}\PY{p}{]} \PY{o}{+} \PY{n}{dismod} \PY{o}{+} \PY{n}{r\PYZus{}offset}\PY{p}{(}\PY{n}{ri\PYZus{}intr}\PY{p}{,} \PY{n}{A0}\PY{p}{,} \PY{n}{logEW}\PY{p}{)}
        \PY{n}{ha} \PY{o}{=} \PY{n}{SED\PYZus{}intr}\PY{p}{[}\PY{l+m+mi}{1}\PY{p}{]} \PY{o}{+} \PY{n}{dismod} \PY{o}{+} \PY{n}{ha\PYZus{}offset}\PY{p}{(}\PY{n}{ri\PYZus{}intr}\PY{p}{,} \PY{n}{A0}\PY{p}{,} \PY{n}{logEW}\PY{p}{)}
        \PY{n}{i} \PY{o}{=} \PY{n}{SED\PYZus{}intr}\PY{p}{[}\PY{l+m+mi}{2}\PY{p}{]} \PY{o}{+} \PY{n}{dismod} \PY{o}{+} \PY{n}{i\PYZus{}offset}\PY{p}{(}\PY{n}{ri\PYZus{}intr}\PY{p}{,} \PY{n}{A0}\PY{p}{,} \PY{n}{logEW}\PY{p}{)}
        \PY{n}{j} \PY{o}{=} \PY{n}{SED\PYZus{}intr}\PY{p}{[}\PY{l+m+mi}{3}\PY{p}{]} \PY{o}{+} \PY{n}{dismod} \PY{o}{+} \PY{l+m+mf}{0.276} \PY{o}{*} \PY{n}{A0}
        \PY{k}{return} \PY{n}{np}\PY{o}{.}\PY{n}{array}\PY{p}{(}\PY{p}{[}\PY{n}{r}\PY{p}{[}\PY{l+m+mi}{0}\PY{p}{]}\PY{p}{,} \PY{n}{ha}\PY{p}{[}\PY{l+m+mi}{0}\PY{p}{]}\PY{p}{,} \PY{n}{i}\PY{p}{[}\PY{l+m+mi}{0}\PY{p}{]}\PY{p}{,} \PY{n}{j}\PY{p}{[}\PY{l+m+mi}{0}\PY{p}{]}\PY{p}{]}\PY{p}{)}

    \PY{c}{\PYZsh{} Likelihood: observed SED}
    \PY{n+nd}{@pymc.stochastic}\PY{p}{(}\PY{n}{observed}\PY{o}{=}\PY{n+nb+bp}{True}\PY{p}{)}
    \PY{k}{def} \PY{n+nf}{SED\PYZus{}observed}\PY{p}{(}\PY{n}{value}\PY{o}{=}\PY{n}{observed\PYZus{}sed}\PY{p}{,} \PY{n}{SED\PYZus{}apparent}\PY{o}{=}\PY{n}{SED\PYZus{}apparent}\PY{p}{)}\PY{p}{:}
        \PY{n}{e\PYZus{}calib} \PY{o}{=} \PY{n}{np}\PY{o}{.}\PY{n}{array}\PY{p}{(}\PY{p}{[}\PY{l+m+mf}{0.1}\PY{p}{,} \PY{l+m+mf}{0.1}\PY{p}{,} \PY{l+m+mf}{0.1}\PY{p}{,} \PY{l+m+mf}{0.1}\PY{p}{]}\PY{p}{)}  \PY{c}{\PYZsh{} Absolute uncertainty term}
        \PY{n}{D2} \PY{o}{=} \PY{n+nb}{sum}\PY{p}{(}\PY{p}{(}\PY{n}{observed\PYZus{}sed} \PY{o}{-} \PY{n}{SED\PYZus{}apparent}\PY{p}{)} \PY{o}{*}\PY{o}{*} \PY{l+m+mi}{2} \PY{o}{/}
                 \PY{p}{(}\PY{n}{e\PYZus{}observed\PYZus{}sed} \PY{o}{*}\PY{o}{*} \PY{l+m+mi}{2} \PY{o}{+} \PY{n}{e\PYZus{}calib} \PY{o}{*}\PY{o}{*} \PY{l+m+mi}{2}\PY{p}{)}\PY{p}{)}
        \PY{n}{logp} \PY{o}{=} \PY{o}{-}\PY{n}{D2} \PY{o}{/} \PY{l+m+mf}{2.0}
        \PY{k}{return} \PY{n}{logp}

    \PY{k}{return} \PY{n+nb}{locals}\PY{p}{(}\PY{p}{)}  \PY{c}{\PYZsh{} Return all model components defined above}


\PY{k}{if} \PY{n}{\PYZus{}\PYZus{}name\PYZus{}\PYZus{}} \PY{o}{==} \PY{l+s}{"}\PY{l+s}{\PYZus{}\PYZus{}main\PYZus{}\PYZus{}}\PY{l+s}{"}\PY{p}{:}
    \PY{l+s+sd}{""" Example code which demonstrates how to obtain the posterior mass """}
    \PY{c}{\PYZsh{} Input: the observed magnitudes and 1-sigma uncertainties}
    \PY{n}{sed\PYZus{}observed} \PY{o}{=} \PY{n}{np}\PY{o}{.}\PY{n}{array}\PY{p}{(}\PY{p}{[}\PY{l+m+mf}{19.41}\PY{p}{,} \PY{l+m+mf}{18.14}\PY{p}{,} \PY{l+m+mf}{17.56}\PY{p}{,} \PY{l+m+mf}{15.44}\PY{p}{]}\PY{p}{)}  \PY{c}{\PYZsh{} r, Ha, i, J}
    \PY{n}{e\PYZus{}sed\PYZus{}observed} \PY{o}{=} \PY{n}{np}\PY{o}{.}\PY{n}{array}\PY{p}{(}\PY{p}{[}\PY{l+m+mf}{0.03}\PY{p}{,} \PY{l+m+mf}{0.03}\PY{p}{,} \PY{l+m+mf}{0.02}\PY{p}{,} \PY{l+m+mf}{0.06}\PY{p}{]}\PY{p}{)}  \PY{c}{\PYZsh{} e\PYZus{}r, e\PYZus{}Ha, e\PYZus{}i, e\PYZus{}J}
    \PY{c}{\PYZsh{} Initialize the model.}
    \PY{n}{mymodel} \PY{o}{=} \PY{n}{make\PYZus{}model}\PY{p}{(}\PY{n}{sed\PYZus{}observed}\PY{p}{,} \PY{n}{e\PYZus{}sed\PYZus{}observed}\PY{p}{)}
    \PY{n}{M} \PY{o}{=} \PY{n}{pymc}\PY{o}{.}\PY{n}{MCMC}\PY{p}{(}\PY{n}{mymodel}\PY{p}{)}
    \PY{c}{\PYZsh{} Demo: run the MCMC sampler and print the expectation value for log(Mass)}
    \PY{n}{M}\PY{o}{.}\PY{n}{sample}\PY{p}{(}\PY{l+m+mi}{50000}\PY{p}{)}
    \PY{n}{samples\PYZus{}logM} \PY{o}{=} \PY{n}{M}\PY{o}{.}\PY{n}{trace}\PY{p}{(}\PY{l+s}{"}\PY{l+s}{logM}\PY{l+s}{"}\PY{p}{)}\PY{p}{[}\PY{p}{:}\PY{p}{]}
    \PY{k}{print} \PY{l+s}{"}\PY{l+s}{logM = }\PY{l+s+si}{\PYZpc{}.2f}\PY{l+s}{ +/-}\PY{l+s+si}{\PYZpc{}.2f}\PY{l+s}{"} \PY{o}{\PYZpc{}} \PY{p}{(}\PY{n}{np}\PY{o}{.}\PY{n}{mean}\PY{p}{(}\PY{n}{samples\PYZus{}logM}\PY{p}{)}\PY{p}{,} \PY{n}{np}\PY{o}{.}\PY{n}{std}\PY{p}{(}\PY{n}{samples\PYZus{}logM}\PY{p}{)}\PY{p}{)}
\end{Verbatim}

\twocolumn


\begin{thebibliography}{}

\bibitem[Bailer-Jones(2009)]{bailer2009} 
Bailer-Jones, C.~A.~L.\ 2009, IAU Symposium, 254, 475

\bibitem[Bailer-Jones(2011)]{bailerjones2011} Bailer-Jones, C.~A.~L.\ 
2011, \mnras, 411, 435 

\bibitem[Barentsen et al.(2011)]{barentsen2011} Barentsen, G., Vink, 
J.~S., Drew, J.~E., et al.\ 2011, \mnras, 415, 103 

\bibitem[Barentsen et al.(2011b)]{barentsen2011b} Barentsen, G., Arlt, 
R., \& Frohlich, H.-E.\ 2011, WGN, Journal of the International Meteor Organization, 39, 126 

\bibitem[Barrado y Navascu{\'e}s \& Mart{\'{\i}}n(2003)]{barrado2003} 
Barrado y Navascu{\'e}s, D., \& Mart{\'{\i}}n, E.~L.\ 2003, \aj, 126, 2997 

\bibitem[Baxter et al.(2009)]{baxter2009} 
Baxter, E.~J., Covey, K.~R., Muench, A.~A., et al.\ 2009, \aj, 138, 963 

\bibitem[Bertout(1989)]{bertout1989}
Bertout, C.\ 1989, \araa, 27, 351

\bibitem[Calvet \& Gullbring(1998)]{calvet1998} 
Calvet, N., \& Gullbring, E.\ 1998, \apj, 509, 802 

\bibitem[Chib \& Greenberg(1995)]{chib1995} 
Chib, S., \& Greenberg, E.\ 1995, American Statistical Journal, 49, 327

\bibitem[Clarke \& Pringle(2006)]{clarke2006} 
Clarke, C.~J., \& Pringle, J.~E.\ 2006, \mnras, 370, L10 

\bibitem[Corradi et al.(2008)]{corradi2008} 
Corradi, R.~L.~M., Rodr{\'{\i}}guez-Flores, E.~R., Mampaso, A., et al.\ 2008, \aap, 480, 409 

\bibitem[Dahm \& Simon(2005)]{dahm2005} Dahm, S.~E., \& Simon, T.\ 2005, \aj, 129, 829 

\bibitem[Dahm et al.(2007)]{dahm2007} 
Dahm, S.~E., Simon, T., Proszkow, E.~M., \& Patten, B.~M.\ 2007, \aj, 134, 999

\bibitem[Dahm(2008)]{handbook2264} 
Dahm, S.~E.\ 2008, Handbook of Star Forming Regions, Volume I, 966 

\bibitem[De Marchi et al.(2010)]{demarchi2010} 
De Marchi, G., Panagia, N., \& Romaniello, M.\ 2010, \apj, 715, 1 

\bibitem[Drew et al.(2005)]{drew2005} 
Drew, J.~E., et al.\ 2005, \mnras, 362, 753
guilar, A., Wang, J., \& Garmire, G.~P.\ 2009, \apj, 699, 1454 

\bibitem[Drew et al.(2008)]{drew2008} 
Drew, J.~E., Greimel, R., Irwin, M.~J., \& Sale, S.~E.\ 2008, \mnras, 386, 1761 

\bibitem[Espaillat et al.(2012)]{espaillat2012} 
Espaillat, C., Ingleby, L., Hern{\'a}ndez, J., et al.\ 2012, \apj, 747, 103 

\bibitem[Evans et al.(2009)]{evans2009} 
Evans, N.~J., et al.\ 2009, \apjs, 181, 321 

\bibitem[Fang et al.(2009)]{fang2009} 
Fang, M., van Boekel, R., Wang, W., Carmona, A., Sicilia-Aguilar, A., \& Henning, T.\ 2009, \aap, 504, 461 

\bibitem[Fedele et al.(2010)]{fedele2010} 
Fedele, D., van den Ancker, M.~E., Henning, T., Jayawardhana, R., \& Oliveira, J.~M.\ 2010, \aap, 510, A72 

\bibitem[Fischer et al.(2011)]{fischer2011} 
Fischer, W., Edwards, S., Hillenbrand, L., \& Kwan, J.\ 2011, \apj, 730, 73 

\bibitem[Flaccomio et al.(2006)]{flaccomio2006} 
Flaccomio, E., Micela, G., \& Sciortino, S.\ 2006, \aap, 455, 903 

\bibitem[Ford(2005)]{ford2005} 
Ford, E.~B.\ 2005, \aj, 129, 1706 

\bibitem[Foreman-Mackey et al.(2012)]{foreman2012} 
Foreman-Mackey, D., Hogg, D.~W., Lang, D., \& Goodman, J.\ 2012, arXiv:1202.3665 

\bibitem[Gennaro et al.(2012)]{gennaro2012} Gennaro, M., Prada 
Moroni, P.~G., \& Tognelli, E.\ 2012, \mnras, 420, 986 

\bibitem[Gonz{\'a}lez-Solares et al.(2008)]{idr} 
Gonz{\'a}lez-Solares, E.~A., et al.\ 2008, \mnras, 388, 89 

\bibitem[Gregory(2005)]{gregory2005} 
Gregory, P.~C.\ 2005, Bayesian Logical Data Analysis for the Physical Sciences: A Comparative Approach with `Mathematica' Support.~Edited by P.~C.~Gregory.~ISBN 0 521 84150 X.~Cambridge University Press, Cambridge, UK, 2005.,  

\bibitem[Groot et al.(2009)]{uvex} 
Groot, P.~J., Verbeek, K., Greimel, R., et al.\ 2009, \mnras, 399, 323 

\bibitem[Gullbring et al.(1998)]{gullbring1998} 
Gullbring, E., Hartmann, L., Briceno, C., \& Calvet, N.\ 1998, \apj, 492, 323 

\bibitem[Gullbring et al.(2000)]{gullbring2000} 
Gullbring, E., Calvet, N., Muzerolle, J., \& Hartmann, L.\ 2000, \apj, 544, 927 

\bibitem[Haisch et al.(2001)]{haisch2001} 
Haisch, K.~E., Jr., Lada, E.~A., \& Lada, C.~J.\ 2001, \apjl, 553, L153 

\bibitem[Hartigan et al.(1995)]{hartigan1995} 
Hartigan, P., Edwards, S., \& Ghandour, L.\ 1995, \apj, 452, 736 

\bibitem[Hartmann(2008)]{hartmannbook} 
Hartmann, L.\ 2008, Accretion processes in star formation. 2nd Edition. Cambridge University Press, Cambridge, UK.

\bibitem[Herczeg \& Hillenbrand(2008)]{herczeg2008} 
Herczeg, G.~J., \& Hillenbrand, L.~A.\ 2008, \apj, 681, 594 

\bibitem[Irwin \& Lewis(2001)]{irwin2001} 
Irwin, M., \& Lewis, J.\ 2001, \nar, 45, 105 

\bibitem[Jeffries et al.(2011)]{jeffries2011} Jeffries, R.~D., 
Littlefair, S.~P., Naylor, T., \& Mayne, N.~J.\ 2011, \mnras, 418, 1948 

\bibitem[J{\o}rgensen \& Lindegren(2005)]{jorgensen2005} J{\o}rgensen, B.~R., \& Lindegren, L.\ 2005, \aap, 436, 127 

\bibitem[Kenyon \& Hartmann(1995)]{kenyon1995} Kenyon, S.~J., \& Hartmann, L.\ 1995, \apjs, 101, 117 

\bibitem[Kipping et al.(2012)]{kipping2012} 
Kipping, D.~M., Bakos, G.~{\'A}., Buchhave, L.~A., Nesvorny, D., 
\& Schmitt, A.\ 2012, arXiv:1201.0752 

\bibitem[Kroupa(2001)]{kroupa2001} 
Kroupa, P.\ 2001, \mnras, 322, 231 

\bibitem[Lucas et al.(2008)]{lucas2008} 
Lucas, P.~W., Hoare, M.~G., Longmore, A., et al.\ 2008, \mnras, 391, 136 

\bibitem[MacKay(2003)]{mackay}
MacKay, D., Information Theory, Inference, and Learning Algorithms, Cambridge University Press, 2003

\bibitem[Martin(1997)]{martin1997} 
Martin, E.~L.\ 1997, \aap, 321, 492 

\bibitem[Meyer et al.(1997)]{meyer1997}
Meyer, M.~R., Calvet, N., \& Hillenbrand, L.~A.\ 1997, \aj, 114, 288 

\bibitem[Mohanty et al.(2005)]{mohanty2005}
Mohanty, S., Jayawardhana, R., \& Basri, G.\ 2005, \apj, 626, 498 

\bibitem[Muzerolle et al.(2003)]{muzerolle2003} 
Muzerolle, J., Hillenbrand, L., Calvet, N., Brice{\~n}o, C., \& Hartmann, L.\ 2003, \apj, 592, 266 

\bibitem[Najita et al.(2007)]{najita2007} 
Najita, J.~R., Strom, S.~E., \& Muzerolle, J.\ 2007, \mnras, 378, 369 

\bibitem[Natta et al.(2004)]{natta2004} 
Natta, A., Testi, L., Muzerolle, J., Randich, S., Comer{\'o}n, F., \& Persi, P.\ 2004, \aap, 424, 603 

\bibitem[Natta et al.(2006)]{natta2006} 
Natta, A., Testi, L., \& Randich, S.\ 2006, \aap, 452, 245 

\bibitem[Owen et al.(2011)]{owen2011} 
Owen, J.~E., Ercolano, B., \& Clarke, C.~J.\ 2011, \mnras, 412, 13 

\bibitem[Park et al.(2000)]{park2000} 
Park, B.-G., Sung, H., Bessell, M.~S., \& Kang, Y.~H.\ 2000, \aj, 120, 894 

\bibitem[Patil et al.(2010)]{pymc}
Patil, A., Huard D., Fonnesbeck C.~J.\ 2010, Journal of Statistical Software, 35, 4, pp.1--81 

\bibitem[Pont \& Eyer(2004)]{pont2004} Pont, F., \& Eyer, L.\ 2004, \mnras, 351, 487 

\bibitem[Rebull et al.(2002)]{rebull2002} Rebull, L.~M., Makidon, 
R.~B., Strom, S.~E., et al.\ 2002, \aj, 123, 1528 

\bibitem[Reipurth et al.(2004)]{reipurth2004} Reipurth, B., 
Pettersson, B., Armond, T., Bally, J., 
\& Vaz, L.~P.~R.\ 2004, \aj, 127, 1117

\bibitem[Rizzuto et al.(2011)]{rizzuto2011} 
Rizzuto, A.~C., Ireland, M.~J., \& Robertson, J.~G.\ 2011, \mnras, 416, 3108 

\bibitem[Robitaille et al.(2006)]{robitaille2006} 
Robitaille, T.~P., Whitney, B.~A., Indebetouw, R., Wood, K., \& Denzmore, P.\ 2006, \apjs, 167, 256 

\bibitem[Robitaille et al.(2007)]{robitaille2007} 
Robitaille, T.~P., Whitney, B.~A., Indebetouw, R., \& Wood, K.\ 2007, \apjs, 169, 328 

\bibitem[Russell \& Norvig(2009)]{aibook} 
Russell, S.~J. \& Norvig, P., 2009, Artificial Intelligence: A Modern Approach. Prentice Hall, Upper Saddle River, NJ

\bibitem[Schlegel et al.(1998)]{schlegel} 
Schlegel, D.~J., Finkbeiner, D.~P., \& Davis, M.\ 1998, \apj, 500, 525 

\bibitem[Sicilia-Aguilar et al.(2010)]{sicilia2010} 
Sicilia-Aguilar, A., Henning, T., \& Hartmann, L.~W.\ 2010, \apj, 710, 597 

\bibitem[Siess et al.(2000)]{siess}
Siess, L., Dufour, E., Forestini, M.\ 2000,\ \aap, 358, 593

\bibitem[Skrutskie et al.(2006)]{2mass} 
Skrutskie, M.~F., et al.\ 2006, \aj, 131, 1163 

\bibitem[Spezzi et al.(2012)]{spezzi2012} 
Spezzi, L., de Marchi, G., Panagia, N., Sicilia-Aguilar, A., \& Ercolano, B.\ 2012, \mnras, 421, 78 

\bibitem[Sung et al.(1997)]{sung1997} 
Sung, H., Bessell, M.~S., \& Lee, S.-W.\ 1997, \aj, 114, 2644 

\bibitem[Sung et al.(2004)]{sung2004} 
Sung, H., Bessell, M.~S., \& Chun, M.-Y.\ 2004, \aj, 128, 1684 

\bibitem[Sung et al.(2008)]{sung2008} Sung, H., Bessell, M.~S., 
Chun, M.-Y., Karimov, R., \& Ibrahimov, M.\ 2008, \aj, 135, 441 

\bibitem[Sung et al.(2009)]{sung2009} Sung, H., Stauffer, J.~R., 
\& Bessell, M.~S.\ 2009, \aj, 138, 1116 

\bibitem[Taylor(2005)]{topcat} 
Taylor, M.~B.\ 2005, Astronomical Data Analysis Software and Systems XIV, 347, 29 

\bibitem[Trotta(2008)]{trotta2008} 
Trotta, R.\ 2008, Contemporary Physics, 49, 71 

\bibitem[Walker(1956)]{walker1956} 
Walker, M.~F.\ 1956, \apjs, 2, 365 

\bibitem[Williams \& Cieza(2011)]{williams2011} 
Williams, J.~P., \& Cieza, L.~A.\ 2011, \araa, 49, 67 

\end{thebibliography}
\end{document}